\documentclass[twocolumn]{aastex631}
\usepackage{bm}
\usepackage{comment}
\usepackage{fontawesome}
\usepackage{aas_macros}
\usepackage{tabularx}
\usepackage{multirow}
\usepackage{float}

\newcolumntype{P}[1]{>{\centering\arraybackslash}p{#1}}

\begin{document}

\title{Inferring halo masses with Graph Neural Networks}

\author[0000-0002-0936-4279]{Pablo Villanueva-Domingo}
\email{pablo.villanueva.domingo@gmail.com}
\affiliation{Instituto de F\'isica Corpuscular (IFIC), CSIC-Universitat de Val\`encia, E-46980, Paterna, Spain}

\author[0000-0002-4816-0455]{Francisco Villaescusa-Navarro}
\email{fvillaescusa@flatironinstitute.edu}
\affiliation{Center for Computational Astrophysics, Flatiron Institute, 162 5th Avenue, New York, NY, 10010, USA}
\affiliation{Department of Astrophysical Sciences, Princeton University, Peyton Hall, Princeton NJ 08544, USA}

\author[0000-0001-5769-4945]{Daniel Angl\'es-Alc\'azar}
\affiliation{Department of Physics, University of Connecticut, 196 Auditorium Road, U-3046, Storrs, CT 06269-3046, USA}
\affiliation{Center for Computational Astrophysics, Flatiron Institute, 162 5th Avenue, New York, NY, 10010, USA}

\author[0000-0002-3185-1540]{Shy Genel}
\affiliation{Center for Computational Astrophysics, Flatiron Institute, 162 5th Avenue, New York, NY, 10010, USA}
\affiliation{Columbia Astrophysics Laboratory, Columbia University, New York, NY, 10027, USA}

\author{Federico Marinacci}
\affiliation{Dipartimento di Fisica e Astronomia ‘Augusto Righi’ , Universit\`a di Bologna, via Gobetti 93/2, 40129, Bologna, Italy}

\author{David N. Spergel}
\affiliation{Center for Computational Astrophysics, Flatiron Institute, 162 5th Avenue, New York, NY, 10010, USA}
\affiliation{Department of Astrophysical Sciences, Princeton University, Peyton Hall, Princeton NJ 08544, USA}

\author{Lars Hernquist}
\affiliation{Center for Astrophysics — Harvard \& Smithsonian, 60 Garden St, Cambridge, MA 02138, USA}

\author{Mark Vogelsberger}
\affiliation{Kavli Institute for Astrophysics and Space Research, Department of Physics, MIT, Cambridge, MA 02139, USA}

\author{Romeel Dave}
\affiliation{Institute for Astronomy, University of Edinburgh, Royal Observatory, Edinburgh EH9 3HJ, UK}
\affiliation{Department of Physics \& Astronomy, University of the Western Cape, Cape Town 7535, South Africa}
\affiliation{South African Astronomical Observatories, Observatory, Cape Town 7925, South Africa}

\author{Desika Narayanan}
\affiliation{Department of Astronomy, University of Florida, Gainesville, FL, USA}
\affiliation{University of Florida Informatics Institute, 432 Newell Drive, CISE Bldg E251, Gainesville, FL, USA}

\begin{abstract}
Understanding the halo-galaxy connection is fundamental in order to improve our knowledge on the nature and properties of dark matter. In this work we build a model that infers the mass of a halo given the positions, velocities, stellar masses, and radii of the galaxies it hosts. In order to capture information from correlations among galaxy properties and their phase-space, we use Graph Neural Networks (GNNs), that are designed to work with irregular and sparse data. We train our models on galaxies from more than 2,000 state-of-the-art simulations from the Cosmology and Astrophysics with MachinE Learning Simulations (CAMELS) project. Our model, that accounts for cosmological and astrophysical uncertainties, is able to constrain the masses of the halos with a $\sim$0.2 dex accuracy. Furthermore, a GNN trained on a suite of simulations is able to preserve part of its accuracy when tested on simulations run with a different code that utilizes a distinct subgrid physics model, showing the robustness of our method. The PyTorch Geometric implementation of the GNN is publicly available on \href{https://github.com/PabloVD/HaloGraphNet}{GitHub \faicon{github}}.
\end{abstract}

\section{Introduction}

In 1933 Fritz Zwicky found out that the mass of the Coma cluster should be much larger than the one from its luminous component \citep{1933AcHPh...6..110Z}. That finding pointed out to the existence of an unknown type of non-luminous matter: dark matter. The requirement of unseen matter was later supported by the observation of rotation curves of galaxies \citep{1978ApJ...225L.107R, 1978PhDT.......195B}. Nowadays, there is overwhelming evidence for the existence of dark matter, although we are still ignorant of its fundamental properties \citep{2017FrPhy..12l1201Y}.

We now believe that dark matter is the backbone of the distribution of matter in the Universe: it concentrates in high-density regions called halos, that are connected by thin filaments with intermediate density and surrounded by gigantic regions with low density (voids). It is within halos that gas can cluster, cool, and form stars and galaxies \citep{2015ARA&A..53...51S}. Dark matter halos are therefore the environment where galaxies reside.

Understanding the halo-galaxy connection is fundamental to improve our knowledge on the nature and properties of dark matter. There are two possible directions to take in the halo-galaxy connection. On one hand, given a dark matter halo and its properties and environment, predict the number and distribution of galaxies it hosts. This task is fundamental in order to create galaxy mocks with the correct clustering on all scales needed for forward modeling approaches \citep{2018ARA&A..56..435W}. On the other hand, given a set of galaxies, it may be useful to determine some properties of their host halo such as its mass, spin, and concentration. This task is fundamental to derive cosmological constraints from the abundance of dark matter halos.

There has been an extensive program of weighing the masses of halos and clusters, using a wealth of techniques including gravitational lensing \citep{2015IAUS..311...86M, 2020MNRAS.492.3685H, 2021arXiv210902646H}, rotation curves of galaxies \citep{2001ARA&A..39..137S, 2015PASJ...67...75S}, abundance matching \citep{2010ApJ...717..379B}, Sunyaev-Zeldovich effect \citep{2001ApJ...552....2G}, X-rays observations \citep{2013MNRAS.433.2790L}, velocity dispersion \citep{2013ApJ...772...47S} and kinematics of satellite galaxies \citep{2013MNRAS.428.2407W, 2020ApJ...903..130S} among others; see, e.g. \cite{2015MNRAS.449.1897O} for a comparison of different techniques for galaxy clusters.

All the above techniques do not make use of all available information. For instance, the abundance matching technique only considers the total stellar mass in the system, disregarding information about its clustering state. In this work we attempt to build a model that can use all available information from observations and/or simulations, e.g. phase-space information, stellar masses, galaxy sizes, to infer the mass of the halo hosting the galaxies. For this, we made use of neural networks and their capacity as universal function approximators.

Different machine learning (ML) algorithms have been already used to perform this task. For instance, Convolutional Neural Networks (CNNs) have been already applied in order to predict dynamical galaxy cluster masses \citep{Ntampaka:2018rjt, Ho:2019zap, Ramanah:2020ift, Ramanah:2020ylz,2020MNRAS.499.3445Y,2020ApJ...900..110G,deAndres:2021tjl}, as well as other ML techniques \citep{Ntampaka:2014ypa, Ntampaka:2015tba, Green:2019uup, 2019MNRAS.484.1526A, 2019arXiv191205316H}. Galaxy and subhalo masses can be inferred from different subhalo properties via multilayer perceptrons \citep{Shao:2021qoa} or other ML algorithms \citep{2021arXiv211101185V}, as well as from images with CNNs \citep{2021arXiv211108725D}. Other works have been employed different ML algorithms to predict the mass of a halo from the properties of the halo and galactic group \citep{2019ApJ...881...74M, 2019MNRAS.490.2367C, 2020arXiv201110577L}. These ML-based approaches have been shown to outperform other traditional techniques to infer halo masses. However, some of these works make use of N-body computations plus semianalytical galaxy formation models, rather than more accurate hydrodynamical simulations. Moreover, several of the features employed may not be easily observable, which could complicate their applicability to real data.

Although these approaches make use of global properties of a halo as well as individual features of the galaxies, they do not incorporate explicitly the relationship between galaxies or subhalos, either in the form of clustering in configuration space and/or distribution in phase-space. In this article, we aim at predicting halo masses exploiting the halo-galaxy connection, using a novel method based on \textit{Graph Neural Networks} (GNNs). This type of neural network shares the typical training procedure of other deep learning techniques, but it is applied to data structured in the form of mathematical graphs. To understand their significance, it is useful to compare them to other deep learning frameworks. CNNs are mostly employed with regular data (grids), such as images and 3D grids; CNNs automatically account for traslational invariance. Recurrent neural networks, on their side, are designed to treat sequential data, such as chains of characters in natural language or time series. However, GNNs can be applied when dealing with irregular data, where data points may have arbitrary relations.\footnote{See \cite{2018arXiv180601261B} for a comparison of the different deep learning components and their relational inductive biases.} That is the case of point clouds, as a galaxy catalogue can be regarded. GNNs have been successfully employed in different fields such as chemistry, computer vision, natural language processing, social networks or particle physics \citep{2019arXiv190100596W, 2021arXiv210413478B}. There are already some applications of GNNs in cosmology, for instance in order to perform symbolic regression \citep{2019arXiv190905862C, Cranmer:2020wew}, to predict the redshift of galaxies \citep{Beck2019RefinedRR}, or to allocate resources in an unsupervised way in order to select galaxies \citep{Cranmer:2021pve}. GNNs have also been applied in other physics fields, such as particle physics \citep{Shlomi:2020gdn}, but still represent a novel, promising and mostly unexplored way to extract information from irregular data.

We train our GNNs using galaxies from simulations of the Cosmology and Astrophysics with MachinE Learning Simulations (CAMELS) project \citep{villaescusanavarro2020camels} to extract information from galaxy properties and their phase-space distribution. Since CAMELS contains thousands of state-of-the-art (magneto-)hydrodynamic simulations with different values of the cosmological and astrophysical parameters, our method accounts for uncertainties in cosmology and astrophysics. Furthermore, since CAMELS contains two different suites of hydrodynamic simulations run with two different codes that employ different subgrid physics, we can quantify the robustness of our results to astrophysical uncertainties. Ultimately, we would like to build a tool that can infer the mass of galaxy systems, like our own Milky Way and Andromeda, just from the observed properties of those galaxies and their satellites. Knowing the mass of the host dark matter halos of those system will allow us to make consistency checks within the $\Lambda$CDM model.

The article is structured as follows. We start by reviewing the basics on graphs and GNNs in Sec. \ref{sec:gnn}. In Sec. \ref{sec:methods} we describe the data we use to train our model together with an outline of the training procedure. The main results of this work are presented in Sec. \ref{sec:results}. Some aspects of the interpretability of the GNNs are examined in Sec. \ref{sec:interpret}, followed by a discussion of the main conclusions in Sec. \ref{sec:discussion}.

\section{Graph Neural Networks}
\label{sec:gnn}

In this section we review the basics of graph neural networks, firstly summarizing the fundamentals of graphs, subsequently detailing how to build graphs from the galaxies of halos, and finally introducing how to build a neural network on a graph based on the message passing scheme. We refer the reader to \cite{2021arXiv210413478B, 2018arXiv180601261B, HamiltonBook} for comprehensive references on graph neural networks and geometric deep learning.

\subsection{General concepts on graphs}

We start by discussing some generic concepts of graphs and standard definitions, since some astrophysicists, cosmologists and, ML practitioners may not be familiar with the terminology. A \textit{graph} can be defined as a tuple $\mathcal{G}=(\mathcal{V},\mathcal{E})$, where $\mathcal{V}$ denotes the set of \textit{nodes}, and $\mathcal{E} \subseteq \mathcal{V} \times \mathcal{V}$ the set of \textit{edges}. For two nodes of the graph $i, j \in \mathcal{V}$, there is an edge connecting them if $(i,j) \in \mathcal{E}$. Nodes $i, j$ are thus coined as \textit{neighbors}. The connectivity can be described by the \textit{adjacency matrix} $A_{ij}$, which takes value of 1 if the pair $(i,j)$ is connected by an edge, being $0$ otherwise. A graph is \textit{undirected} when, if $(i,j) \in \mathcal{E}$, then $(j,i) \in \mathcal{E}$ too. A \textit{loop} is an edge which connects a node to itself. We restrict our discussion to \textit{simple} graphs, i.e., undirected graphs without loops, since these are enough for our purposes (although self-loops can be implicitly employed in some GNN architectures).

Given a node $i$, we denote its \textit{feature vector} as $\textbf{x}_i \in \mathbb{R}^{n_{\rm in}}$, encoding the relevant physical information about the node, with $n_{\rm in}$ the number features. The \textit{feature matrix} $\textbf{X} = (\textbf{x}_1, ... , \textbf{x}_{|\mathcal{V}|})^T \in \mathbb{R}^{|\mathcal{V}|\times n_{\rm in}}$, where $|\mathcal{V}|$ is the number of nodes in the graph, comprises the feature vectors of all nodes. The \textit{neighborhood} of the node $i \in \mathcal{V}$, denoted by $\mathcal{N}_i$, includes every node which shares and edge with $i$, and can thus be written as
\begin{equation}
\mathcal{N}_i = \{ j \; | \; A_{ij}=1 \}.
\end{equation}
Note that $\mathcal{N}_i$ does not include the node $i$, since we dismiss loops. Analogously, we can define the set of feature vectors of the neighbors of the node $i$ as
\begin{equation}
    \mathcal{X}_i=\{\textbf{x}_j \; | \; j \in\mathcal{N}_i \}.
\end{equation}
A graph is \textit{complete} if every pair of nodes is connected by an edge, having thus ${|\mathcal{V}| \choose 2} = |\mathcal{V}|(|\mathcal{V}|-1)/2$ edges in total.\footnote{This can be reasoned as the number of pairs that can be chosen from a total of $|\mathcal{V}|$ elements.} The edges set can hence be written as $\mathcal{E}=\mathcal{V} \times \mathcal{V} \setminus \{(i,i) \in \mathcal{V}\}$ (excluding loops). As will be shown later, most of the halos employed here lead to complete graphs.\footnote{Complete graphs can be very relevant in ML applications, since the popular architecture Transformers \citep{2017arXiv170603762V} can be regarded as a particular case of a GNN, a Graph Attention Network \citep{velickovic2018graph}, where the graph is complete (see \citealt{2021arXiv210413478B} for more details).}

\subsection{Halos as graphs}
\label{sec:halosasgraphs}

\begin{table}
	\begin{center}
	\setlength{\tabcolsep}{10pt}    
	\renewcommand{\arraystretch}{1.3}   
		\begin{tabular}{|c|c|}
			\hline
			Symbol & Feature \\
			\hline
			\hline
			 $\textbf{p}$ & 3D comoving position \\ 
			 \hline
			 $v$ & Modulus of relative velocity \\ 
			 \hline
			 $M_*$ & Stellar mass \\
			 \hline
			 $R_*$ & Stellar half-mass radius \\
			 \hline
		\end{tabular}
	\end{center}
	\caption{Summary of galactic properties employed to train the GNNs.}
	\label{table:features}
\end{table}

\begin{figure*}[th!]
\begin{center}
\includegraphics[width=0.9\linewidth]{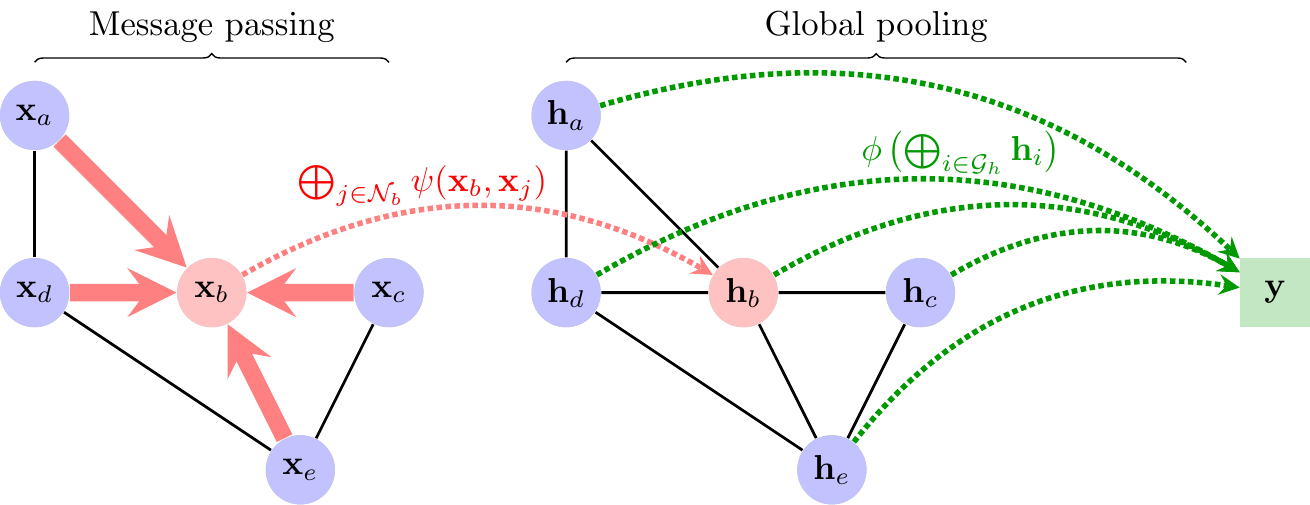}
\caption{Sketch of a GNN with one message passing layer acting on a graph $\mathcal{G}_h$ composed of 5 nodes. During the message passing step, node $b$ receives a message from each of its neighbors $j \in \mathcal{N}_b$, aggregating and updating its features to the hidden vector $\textbf{h}_b$ (red). Once this step is done over all nodes, a global pooling step is performed, aggregating the hidden node features over all vertices of the graph $\mathcal{G}_h$, which leads to the global target $\textbf{y}$ (green). Figure based on \url{https://github.com/PetarV-/TikZ}.}
\label{fig:messagepassing}
\end{center}
\end{figure*}

The general terminology discussed above allows us to set up our problem. Consider a halo $h$ with mass $M_h$ which hosts several  galaxies. We build a graph, denoted by $\mathcal{G}_h$, by considering all the (central and satellite) galaxies of the halo as the nodes of the graph. The feature vector of a node $i$, $\textbf{x}_i$, will include information about the corresponding galaxy, namely the 3D comoving position, $\textbf{p}$, the stellar mass, $M_*$, the modulus of the velocity, $v$, and the stellar half-mass radius (i.e., the comoving radius containing half of the stellar mass in the galaxy), $R_*$. These galactic properties are listed on Table \ref{table:features}. Positions are expressed with respect to the center of the halo (chosen as the point with minimum gravitational potential energy), while velocities are written relative to the center of mass frame of all particles.\footnote{We have checked that using only distances rather than 3D positions significantly worsens the performance of the GNN, since it does not provides a complete description of the spatial distribution of the system, which can be relevant for the halo/galaxy connection.} The mass of the halo is taken as the total mass of all components (dark matter, gas, stars, and black holes) within a virial radius enclosing a density 200 times the critical density, usually known as $M_{200c}$.

Unlike other graph applications, such as, e.g., social networks, we are dealing with distributions of particles not physically connected among them (sometimes known as \textit{point clouds} in graph contexts), and thus the edges are not predetermined. There is not an unique way to build the edges between the nodes. Since gravity is a long-range force, every galaxy is affected by the rest, and all the nodes should be connected in some way. However, gravitational effects, such as tidal forces, are stronger for shorter distances. Hence, one can expect that the dynamics and properties of a given galaxy will be more affected by nearby companions rather than by distant ones. Different approaches can be followed to connect the graphs and build the neighborhood of each node. In our case, we create edges from a node taking into account the neighbors within a fixed certain radius, which is taken as a hyperparameter. Its specific value is given from the optimization outlined in Sec. \ref{sec:training}. A similar procedure could be considering the $k$ nearest neighbors for each node. However, this approach does not properly reflect the effect of clustering. It may miss some neighbors in a clustered region which should be connected, and include others much farther away, which in principle should have less influence in the node. We have checked that performance slightly worsens using this approach instead of considering nodes within a certain distance. It is worth it to remark that we only consider galaxies that are part of the same halo, excluding thus interlopers and explicit relations between halos.

Among the available halos in the CAMELS catalogues, we work only with those containing more than one galaxy, excluding therefore halos without satellites. The halos excluded,  typically very low mass halos, form trivial single-node graphs, where the advantages of graph structured data with edge connections between nodes is absent and therefore are not very interesting for our purposes here. We have explicitly checked that by including those halos without satellites our results worsen. Note also that this fact limits the application of our method to real observations to halos with more than one known satellite galaxy.

\subsection{Message passing}

\label{sec:messagepassing}

GNNs employ the so-called neighborhood aggregation or \textit{message passing} scheme.\footnote{There are other similar frameworks, such as the Graph Network formalism \citep{2018arXiv180601261B}, which is mostly equivalent to our message passing scheme.} Each node $i$ receives a \textit{message} from each of its neighbors $j$, appending its information to its own features. This makes it possible to create a \textit{hidden} feature vector $\textbf{h}_i$, which updates the node and contains aggregated data from its neighborhood. From the hidden vectors, our GNN is aimed at inferring the mass of the halo, which is a global property of the graph. The architecture is thus composed of two main steps: 1) update nodes via message passing, and 2) extract global information of the graph. We designate as \textit{Graph Layer}, GL, to a generic message passing layer, which can be written as 
\begin{equation}
 \textbf{h}_i = {\rm GL}(\textbf{x}_i, \mathcal{X}_i) = \bigoplus_{j \in \mathcal{N}_i} \psi(\textbf{x}_i,\textbf{x}_j),
 \label{eq:messagepassing}
\end{equation}
where $\textbf{h}_i$ is the output feature vector, $\bigoplus$ denote a differentiable, permutation invariant \textit{aggregation function}, such as the maximum, the mean or the sum, while $\psi$ is the \textit{message function}, a differentiable function, like a Multi Layer Perceptron (MLP), which depends upon the feature vector of the node, $\textbf{x}_i$, as well as on those from its neighbors, $\textbf{x}_j$, for $j \in \mathcal{N}_i$. The distinct ways to build $\psi$ lead to different graph layers.\footnote{One could also apply an additional MLP $\xi$ to $\textbf{h}_i$ and its original feature vector, $\textbf{h}_i' = \xi([\textbf{x}_i, \textbf{h}_i])$, where the square brackets indicate array concatenation, although we dismiss this step for the sake of simplicity, and since we apply several successive graph layers, which leads to a similar effect.}

The function $\psi$ maps $\psi: \mathbb{R}^{n_{\rm in}} \rightarrow \mathbb{R}^{n_{\rm out}}$, with $n_{\rm in}$ and $n_{\rm out}$ the number of initial and output features, respectively. As shown in Sec. \ref{sec:architecture}, $n_{\rm in}$ may not correspond to the number of features considered of each node. A common advantage of GNNs is that they exploit locality of data, since closer nodes may present stronger relations than those farther away. Here,  locality is enforced by assuming that the functions $\psi$ are the same for all nodes and graphs.\footnote{Similarly, CNNs in, e.g., computer vision tasks guarantee locality by employing the same kernels over all different images.}

The final GNN incorporates a last step to infer the global graph target, and thus can be of the form
\begin{equation}
\textbf{y} = \phi\left( \bigoplus_{i \in \mathcal{G}_h} \textbf{h}_i \right),
 \label{eq:gnn}
\end{equation}
where $\phi$ is a differentiable function, chosen as a MLP with three hidden layers with 300 channels separated by ReLu activation functions, and $\textbf{y}$ is the output of the network. In our case, since we perform likelihood-free inference to estimate the mean and standard deviation of the posterior of the mass of the halo, the target is given by a vector of two components $\textbf{y}=[ y,\sigma ]$, containing the mean prediction $y$ and its expected standard deviation, $\sigma$. The targeted quantity is the logarithm of the mass of the halo, $y=log_{10}\left[M_h/(M_\odot/h)\right]$. The likelihood-free approach to estimate the posterior mean and standard deviation is detailed in Sec. \ref{sec:training}. The equation above can be easily modified to include some global graph quantities $u$, which may help to train the network, such as the number of nodes or the total stellar mass, writing $\textbf{y} = \phi\left( [\bigoplus_{i \in \mathcal{G}_h} \textbf{h}_i, u] \right)$. We include these global features in our graphs, although we have checked that removing them does not leave a significant impact in the results. Figure \ref{fig:messagepassing} shows an sketch of this basic architecture, illustrating the message passing scheme.

Permutation equivariance and invariance are key concepts in GNNs. We say that a function acting on the node feature matrix $f(\textbf{X})$ is permutation invariant if for every permutation matrix $\textbf{P}$, one has $f(\textbf{P}\textbf{X})=f(\textbf{X})$, leaving thus unchanged the output. On the other hand, $f$ is permutation equivariant if it transforms as $f(\textbf{P}\textbf{X})=\textbf{P}f(\textbf{X})$.\footnote{See \cite{2021arXiv210413478B} for more details on invariances on graphs.} A message passing layer ${\rm GL}(\textbf{x}_i, \mathcal{X}_i)$ should be permutation equivariant, since a reordering of the input nodes produces the same permutation in the outputs, although the output feature space can be different. However, any global quantity of the graph such as the final output of the network, the halo mass, must be permutation invariant, since its value should not depend on the ordering of the nodes. The final aggregation step, $\bigoplus_{i \in \mathcal{G}_h} {\rm GL}(\textbf{x}_i, \mathcal{X}_i)$, can be regarded as a function $f(\textbf{X})$ which fulfills by construction permutation invariance, $f: \mathbb{R}^{|\mathcal{V}|\times n_{\rm out}} \rightarrow \mathbb{R}^{n_{\rm out}}$. In this case, the aggregation $\bigoplus$ may be termed as a global pooling layer, since it reduces dimensionality. Note that the aggregation method here does not have to be the same as in the message passing layer. Furthermore, it is possible to employ different aggregation operators, such as maximum, sum, and mean, and concatenate them, which is the approach used in this work. The last MLP $\phi$ makes use of $n_{\rm out}$ global features of the graph in order to extract a global property vector $\textbf{y}$, the halo mass and its expected standard deviation in our case, $\phi: \mathbb{R}^{n_{\rm out}} \rightarrow \mathbb{R}^2$. Finally, we can apply multiple successive GNN layers, ${\rm GL}({\rm GL}(\textbf{x}_i, \mathcal{X}_i), \mathcal{H}_i)$, where $\mathcal{H}_i = \{\textbf{h}_j \; | \; j \in\mathcal{N}_i \}$ denotes the updated feature vector of the neighbors. In such a case, the node after $k$ updates will encode information from the nodes of its $k$-hop neighborhood. We take the number of message passing layers as an optimizable hyperparameter, although a single GNN layer is sufficient to provide the best results, as shown in Sec. \ref{sec:training}.

\subsection{Graph layer architecture}
\label{sec:architecture}

So far we have discussed the general shape of the GNN, but the explicit design of the message passing layer remains undetermined. There are many possible ways to construct the specific architecture of the GNN layer, which is defined by the message function $\psi$ and the aggregation scheme $\bigoplus$. For instance, one could not take into account neighbors to update each node, and employ only the information of the node $\textbf{x}_i$ to extract the hidden information $\textbf{h}_i$. These types of architectures, a subset of GNNs, are known as DeepSets \citep{2017arXiv170306114Z}. No aggregation function is thus needed to update the node features, and the hidden layer can just be written as $\textbf{h}_i = \psi(\textbf{x}_i)$. 

However, to fully exploit the relations among the nodes, it is useful to incorporate neighborhood and edge information actually performing message passing. Here we assume an architecture based on the edge convolutional layer, coined as EdgeNet \citep{wang2019dynamic}, where for each neighbor $j$ of the node $i$, the relative vectors $\textbf{x}_i-\textbf{x}_j$ are concatenated to the feature vector $\textbf{x}_j$. The aggregation function is chosen as the maximum, writing thus the hidden layer as
\begin{equation}
 \textbf{h}_i = \max_{j \in \mathcal{N}_i} \; \psi([\textbf{x}_i,\textbf{x}_i-\textbf{x}_j]) ,
 \label{eq:edgenet}
\end{equation}
where the square brackets indicate array concatenation. In this case, the initial input number is given by $n_{\rm in}=2n_{\rm feat}$, where $n_{\rm feat}$ is the number of features. The differentiable function $\psi$ is taken as a MLP with three hidden layers, with 300, 300 and 100 hidden channels, separated by ReLu activation functions. We have checked that other choices of the number of channels do not improve the preditions of the net.

There are other popular architectures, such as PointNet \citep{qi2017pointnet,qi2017pointnet++}, which employs only the relative spatial positions for the message passing, $\textbf{p}_i-\textbf{p}_j$, rather than the full feature vector, or the Graph Convolutional Network \citep{kipf2017semisupervised}, where the neighbor information is simply incorporated by taking a linear combination of the features and summing over all the neighbors, imitating a convolution operation. We have checked that the EdgeNet (Eq. \ref{eq:edgenet}) outperforms the other architectures mentioned above via a hyperparameter optimization procedure, as commented in Sec. \ref{sec:training}. See, e.g., \cite{2019arXiv190100596W, 2018arXiv181208434Z} for a discussion of other different GNN architectures.\footnote{See also \url{https://github.com/thunlp/GNNPapers} for a comprehensive list of GNN references.}

\section{Methods}

\label{sec:methods}

In this section, we specify the details regarding the data employed and the training of the network. 

\subsection{The CAMELS simulations}
\label{sec:camels}

The CAMELS project \citep{villaescusanavarro2020camels, 2022arXiv220101300V} comprises a set of state-of-the-art hydrodynamic and N-body simulations, specially suited and designed to train and test ML algorithms. They include thousands of realizations varying two cosmological parameters, namely the matter density parameter $\Omega_m$, and the variance of the linear field on 8 Mpc$/h$ at $z=0$, $\sigma_8$, plus four astrophysical parameters controlling the efficiency of supernovae and active galactic nuclei (AGN) feedback. The rest of cosmological parameters are kept fixed to standard values in flat cosmologies with cosmological constant: the baryonic fraction $\Omega_b=0.049$, the reduced Hubble constant $h=0.6711$, the tilt of the primordial power spectrum $n_s=0.9624$, the equation of state of the dark energy (corresponding to a cosmological constant) $w=-1$, and the total neutrino mass $M_\nu=0$ eV. Each simulation follows the evolution from $z=127$ to $z=0$ of $256^3$ DM particles and $256^3$ gas resolution elements within a box of size periodic volume of size $25$ Mpc$/h$. The DM particle mass resolution is $\sim 10^8 M_\odot/h$. A collection of different cosmological fields in form of 2D maps and 3D grids from the CAMELS simulations is available as the CAMELS Multifield Dataset (CMD)\footnote{\url{https://camels-multifield-dataset.readthedocs.io}}, intended to be a standard cosmological dataset for ML applications \citep{2021arXiv210910915V}, while the full dataset has been recently made publicly available \citep{2022arXiv220101300V}.

The CAMELS project includes two suites of simulations, with different astrophysics modeling and subgrid physics. On the one hand, a suite of magneto-hydrodynamic simulations with the subgrid physics models of IllustrisTNG \citep{2017MNRAS.465.3291W, 2018MNRAS.473.4077P, 2019ComAC...6....2N}, performed with the code Arepo\footnote{\url{https://arepo-code.org/}} \citep[][see also \citealt{2018MNRAS.475..676S,2018MNRAS.480.5113M,2018MNRAS.477.1206N} for more details]{Weinberger:2019tbd}. Its galaxy formation model is based on the predecessor Illustris \citep{2013MNRAS.436.3031V, 2014Natur.509..177V}. On the other hand, another suite employs the SIMBA subgrid physics model \citep{Dave:2019yyq}, making use of the code
GIZMO\footnote{\url{http://www.tapir.caltech.edu/~phopkins/Site/GIZMO.html}} \citep{Hopkins:2014qka}. SIMBA is built on its precursor MUFASA \citep{2016MNRAS.462.3265D} with the addition of supermassive black hole growth and feedback \citep{2017MNRAS.464.2840A}. The hydrodynamics simulations are accompanied by N-body counterparts, run with the GADGET-III code \citep{Springel:2005mi}.\footnote{See, e.g., \cite{2020NatRP...2...42V} for a comparison of different models of galaxy formation in cosmological simulations.}

Within each simulation suite, there are different sets, according to the configuration of astrophysical and cosmological parameters. The CV simulations (standing for Cosmic Variance) include 27 simulations sharing their astrophysical and cosmological parameters, fixed to standard values ($\Omega_m = 0.3$ and $\sigma_8=0.8$), and varying only the random seed to generate the initial conditions. The LH set (standing for Latin-Hypercube) consists of 1,000 simulations, varying all the astrophysical and cosmological parameters (together with the random seed) along a latin-hypercube.\footnote{Besides CV and LH, other simulation sets are also included in CAMELS, but not employed in this work.} We make use of both sets for training our networks, in order to check whether the architectures considered are robust enough for different cosmologies and baryonic feedback parameters. The halos and subhalos are identified using the \textsc{SUBFIND} algorithm \citep{Subfind}. We define galaxies as subhalos that contain more than 20 star particles. Although the CAMELS suite contains data for several redshifts, only simulations at $z=0$ are considered in this work. We refer the reader to the CAMELS webpage\footnote{\url{https://camels.readthedocs.io}} and \cite{villaescusanavarro2020camels} for further details.

\begin{figure}[t!]
\begin{center}
\includegraphics[width=0.99\linewidth]{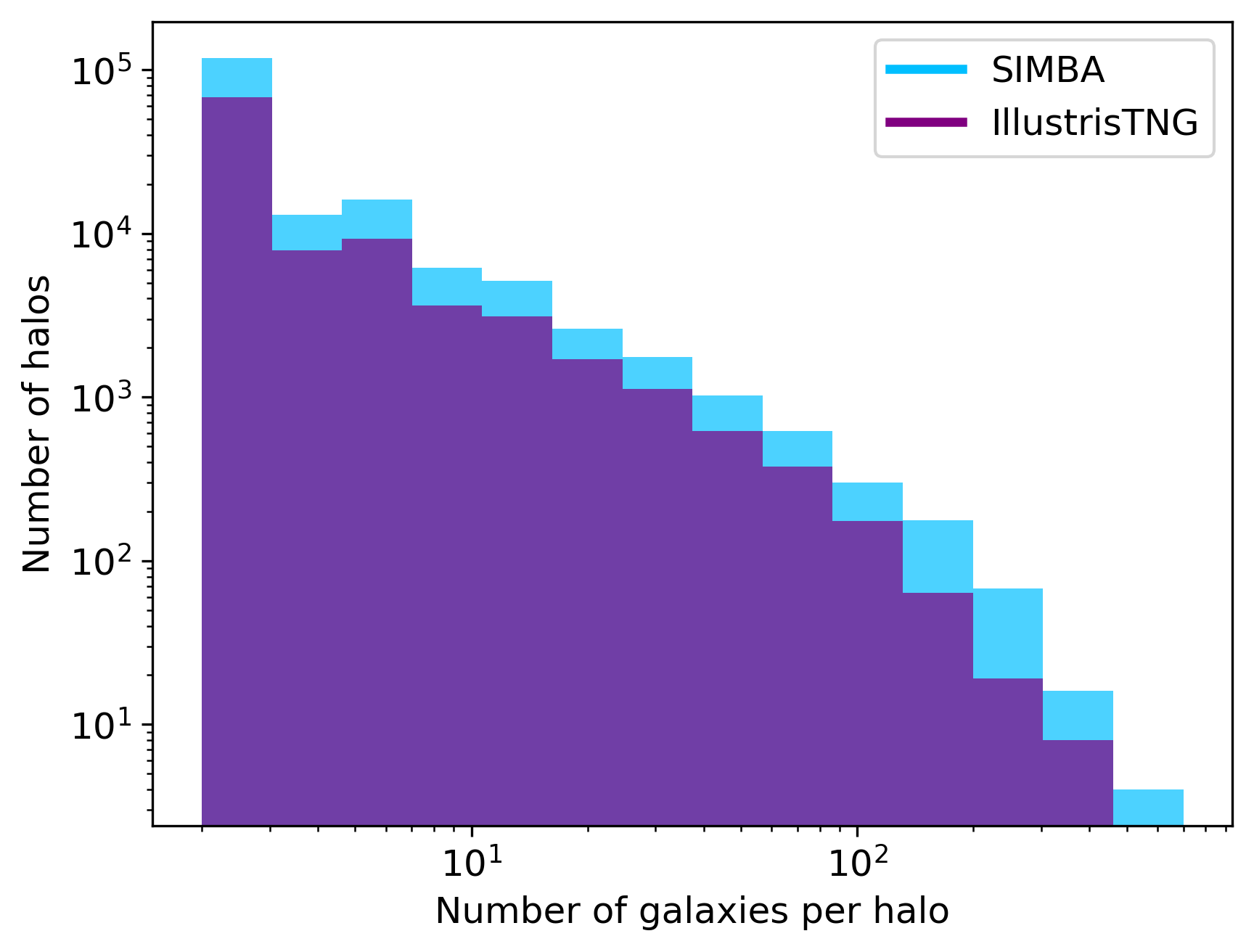}
\caption{Number of halos sorted by the number of galaxies that they host in the IllustrisTNG (blue) and SIMBA (purple) suites, for the LH sets. On average, the halos of the SIMBA simulations contain more galaxies than the halos of the IllustrisTNG simulations, reflecting the large differences between the two simulation suites given their distinct subgrid physics models.}
\label{fig:histogram}
\end{center}
\end{figure}

As an example of the differences between IllustrisTNG and SIMBA, Fig. \ref{fig:histogram} shows the number of halos as a function of the number of galaxies the halos contain for the simulations of the LH set. While most of the halos in the simulations only host a few galaxies, there are a number of them that contain hundreds of satellites. Note that for a fixed halo mass, the SIMBA simulations contain more galaxies than the IllustrisTNG simulations. In general, SIMBA simulations usually contain many more galaxies than their IllustrisTNG counterpart, being an average of $\sim 1200$ and $\sim 700$ per simulation respectively, both at $z=0$. Furthermore, SIMBA simulations tend to be more stochastic, and usually can produce more extreme scenarios (with larger effects of the baryonic feedback effects) than IllustrisTNG. These facts reflect the inherent differences between the two codes/subgrids models. We refer the reader to \cite{villaescusanavarro2020camels} for a more detailed comparison between different observables of both suites, such as clustering or halo population, and to \cite{2022arXiv220101300V, 2022arXiv220413713V} for more information on the distribution and correlations between the different galactic features of the CAMELS catalogues.

\subsection{Training procedure}
\label{sec:training}

We construct our dataset, the collection of graphs, based on the procedure outlined in Sec. \ref{sec:halosasgraphs}. We restrict our analysis to halos with more than one galaxy, as mentioned above. The models are trained on simulations of the CV and LH sets separately, where in each case we split the dataset into training (70\%), validation (15\%) and testing (15\%) sets. We employ an Adam optimizer \citep{kingma2014method} and L2 regularization. The batch size is set to 128 and the number of epochs limited to 150. The GNN architecture is implemented following the prescription outlined in Sec. \ref{sec:messagepassing}, concatenating one or several message passing layers and appending at the end a global pooling and MLP to infer the halo mass.

We perform a hyperparameter optimization in order to get the best values for the hyperparameters, following a bayesian model-based optimization procedure with the Tree Parzen Estimator \citep[TPE,][]{NIPS2011_86e8f7ab}, making use of the Python package Optuna\footnote{\url{https://optuna.readthedocs.io}} \citep{2019arXiv190710902A}. The hyperparameters considered for this optimization are the learning rate, the weight decay, the number of message passing layers, the distance to define neighborhoods, and the specific architecture (among those defined in Sec. \ref{sec:architecture}). We perform at least 75 trials for each suite and set, where each trial is a specific choice of the values of the hyperparameters. The optimization procedure leads to different values for each simulation suite and set, with learning rates ranging between $10^{-5}$ and $6 \times 10^{-4}$, weight decay values between $10^{-8}$ and $10^{-7}$, and the neighborhood distance  between 2 and 20 Mpc/h. One unique message passing layer (i.e., one update of the hidden features of the nodes from Eq. \ref{eq:messagepassing}) and the EdgeNet architecture (among those discussed in Sec. \ref{sec:architecture}) are the optimal choices for all cases. The large values of the neighborhood radii obtained imply that most of the graphs created are complete, around 98\% for both the CV and LH sets. That means that most of the galaxies are connected with each other within the halo, and most nodes can access to information about each other companion of the graph at one hop. 

\begin{figure*}[ht!]
\begin{center}
\includegraphics[width=0.49\linewidth]{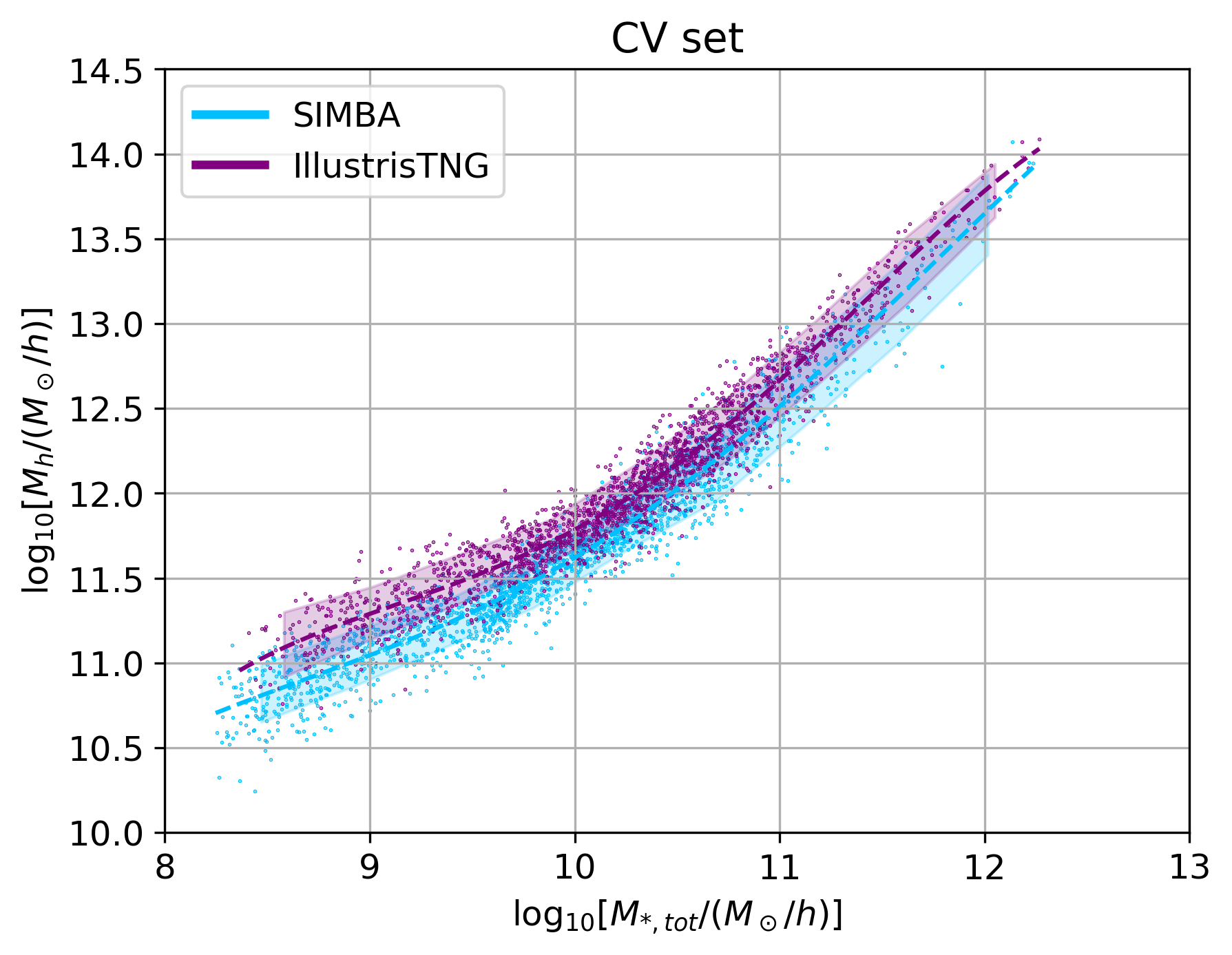}
\includegraphics[width=0.49\linewidth]{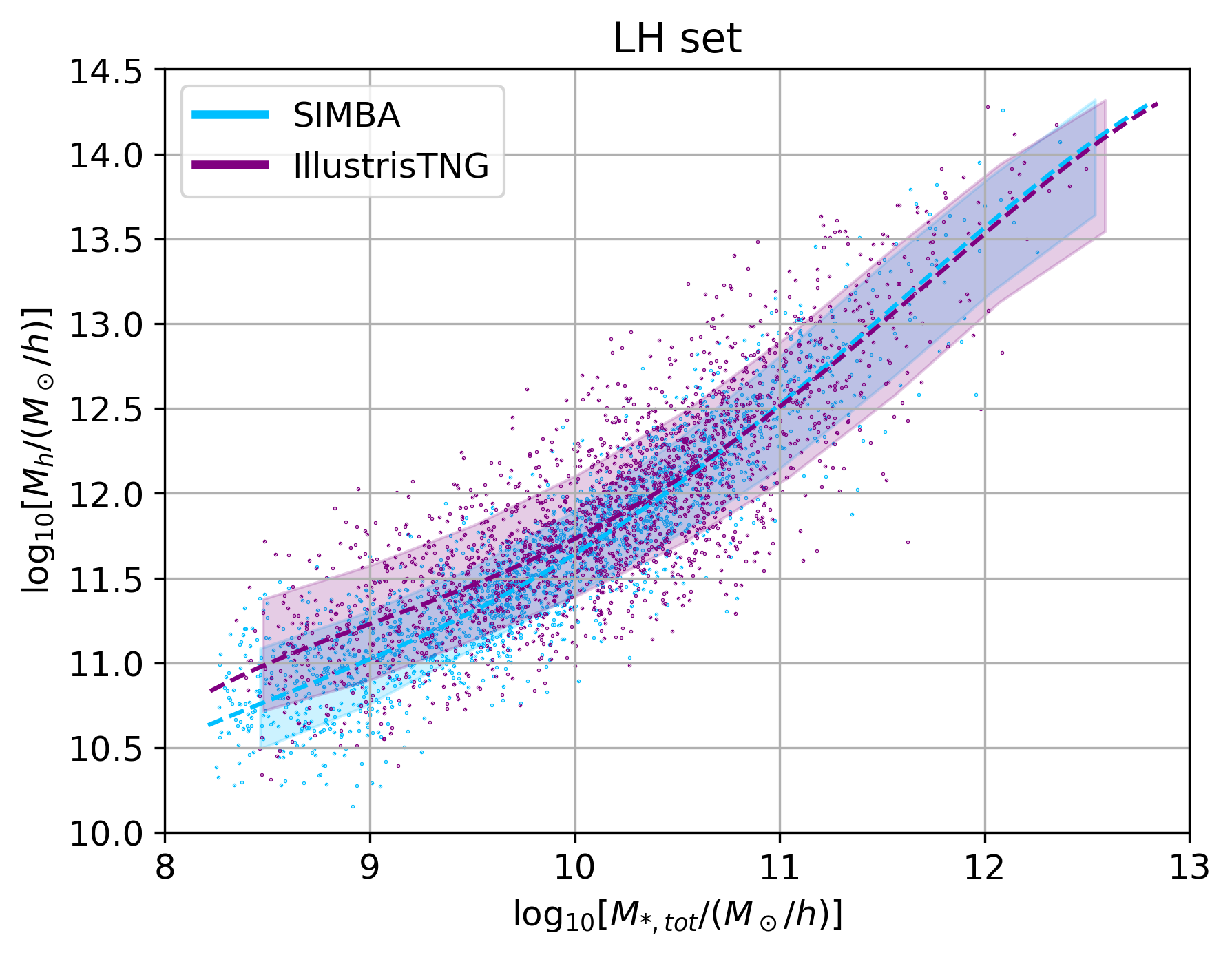}
\caption{Halo masses with respect to the total stellar mass within, both for SIMBA and IllustrisTNG suites, for samples of 2000 halos from the CV (left) and LH (right) sets. Shaded areas denote standard deviation of points while dashed lines correspond to polynomial fits. The LH set covers a large astrophysical and cosmological parameter space, leading to a broader scatter in the masses with respect to the CV set, where parameters are fixed. While such fits can lead to accurate predictions of the halo mass for the CV set, it worsens in the LH set, given the larger dispersion.}
\label{fig:scatplot}
\end{center}
\end{figure*}

We follow a likelihood-free bayesian inference approach to sample the expected standard deviation of the outputs. This procedure allows us to reproduce some properties of the posterior (its second centered moment in this case) without the need of a likelihood \citep{Jeffrey:2020xve}. We follow \citet{moment_networks} and design our model to output two quantities: the mean and standard deviation of the halo mass posterior. The loss function needed to achieve that is given by  $\mathcal{L}=\mathcal{L}_1+\mathcal{L}_2$, where
\begin{equation}
    \mathcal{L}_1=\log\left[\sum_{i\in{\rm batch}}\left(y_{{\rm truth}, i} - y_{{\rm infer}, i}\right)^2\right]
\end{equation}
and
\begin{equation}
    \mathcal{L}_2=\log\left[\sum_{i\in{\rm batch}}\left(\left(y_{{\rm truth}, i} - y_{{\rm infer}, i}\right)^2 - \sigma_{i}^2 \right)^2\right]
    \label{eq:loss2}
\end{equation}
Note that we have included \textit{log} functions to make sure that the two contributions are similar. For instance, if the errors are much smaller than the mean, the loss will be dominated by the mean and the errors may not be accurately computed \citep[see][for further details]{2021arXiv210910915V}.

It is worth emphasizing the symmetries fulfilled by the GNNs. By construction, GNNs are permutation invariant, guaranteed by neighborhood aggregation as discussed in Sec. \ref{sec:messagepassing}. Furthermore, given that graphs are written in the center of mass rest frame, our framework is also translational invariant, since any displacement of the galaxy group would not alter the relative coordinates with respect to the center, and hence the graph would be the same. Moreover, for our special case of point clouds, our GNNs should be rotationally invariant, since an arbitrary rotation of all galaxies around the center of the halo should not change the global halo properties. We try to enforce this symmetry by performing random rotations on each graph at every training epoch, as a data augmentation procedure. This practice also helps in alleviating overfitting, given that the network becomes robust when tested in systems with arbitrary rotations. Overfitting also becomes absent in our training procedure with the proper selection of the hyperparameters.

We write and train the models making use of PyTorch Geometric\footnote{\url{https://pytorch-geometric.readthedocs.io}} \citep{Fey_Fast_Graph_Representation_2019}. Our implementation of the GNNs, {\tt HaloGraphNet}, is publicly available on \href{https://github.com/PabloVD/HaloGraphNet}{GitHub \faicon{github}}\footnote{\url{https://github.com/PabloVD/HaloGraphNet}} \citep{HaloGraphNet}.

\section{Results}
\label{sec:results}

In this section, the main results of the work are discussed. We first introduce a benchmark model to estimate halo masses from stellar masses, and next we detail the results of training and testing the GNNs in the different CAMELS simulation sets and suites considered, examining also their robustness over different astrophysical modeling.

\subsection{Predictions from stellar masses}
\label{sec:benchmark}

Before starting to discuss the accuracy of the GNN predictions, it may be useful to set a benchmark model to predict halo masses, to test the worthiness and degree of improvement of using GNNs. A traditional approach exploits the relation between stellar and halo mass, based on abundance matching \citep{2010ApJ...717..379B, 2018ARA&A..56..435W}. Figure \ref{fig:scatplot} shows the mass of a halo $M_h$ as a function of its total stellar mass $M_{*,tot}=\sum_i M_{*,i}$ (summing over all stellar particles in the central and satellite galaxies within the halo) in the IllustrisTNG and SIMBA suites, for the CV (left) and the LH (right) sets.\footnote{Note that other works consider the stellar mass of the central galaxy rather than the total one, which is employed here with the aim to take into account the contribution of the satellites.} One can notice a clear correlation between stellar and halo masses, although with a large scatter. This is specially noteworthy in the LH set, where multiple values of the cosmological and astrophysical parameters are considered, leading to completely different outcomes. Shaded areas denote the standard deviation of the points, which is around $\sim 0.2$ dex in the CV case, but grows up to $\sim 0.3-0.4$ dex in the LH set.

Taking advantage of the expressed correlation, we can build a naive estimator of the halo mass based only on the total stellar mass of galaxies. A simple 4th degree polynomial fit is shown in dashed lines in Fig. \ref{fig:scatplot}. This can be regarded as a benchmark model for comparing with our forthcoming models based on GNNs, in order to further evaluate their strength. To check its accuracy for predicting $y=\log_{10}\left[M_h/(M_\odot/h)\right]$, we employ the mean relative error $\epsilon$, 
\begin{equation}
    \epsilon = \frac{1}{N} \sum_i^N \frac{|y_{{\rm truth}, i} - y_{{\rm infer}, i}|}{y_{{\rm truth}, i}},
\end{equation}
with $N$ the number of test halos, as well as the correlation coefficient (or coefficient of determination) $R^2$, defined in the usual way as
\begin{equation}
    R^2 = 1 - \frac{\sum_i^N (y_{{\rm truth}, i} - y_{{\rm infer}, i})^2}{\sum_i^N (y_{{\rm truth}, i} - \overline{y}_{{\rm truth}})^2},
\end{equation}
with $\overline{y}_{{\rm truth}}$ the mean of true values. This naive estimator gives fairly accurate results in the CV set, with relative errors around 1\% and a linear correlation coefficient of $R^2\simeq 0.94$. However, when an analogous fit is attempted in the LH case, the relative error worsens down to $\sim 1.7$\% ($\sim 2.4$\%) for SIMBA (IllustrisTNG), with $R^2\simeq 0.84$ ($R^2\simeq 0.67$). Note that these errors are for the logarithm of the mass, $y$, rather than for the halo mass itself, which correspond to relative errors in the mass between $\sim 50-120 \%$. This fact illustrates the non-trivial dependence of the halo and stellar masses on the different astrophysical and cosmological scenarios. Notice that the LH set contains some extreme scenarios which may be unlikely to describe the real universe. In any case, the most appropriate astrophysical parameters to describe our cosmos are still unclear, and thus, models able to robustly marginalize over baryonic feedback effects are required. Thus, a prediction of the halo mass from only the stellar mass when a broad range of astrophysical models are considered seems to be quite inaccurate. In the following, we shall show how a GNN is able to overcome this difficulty by considering further features and taking advantage of the graph structure of halos.

\subsection{Inferring halo masses with GNNs}

\begin{figure*}[th!]
\begin{center}
\includegraphics[width=0.49\linewidth]{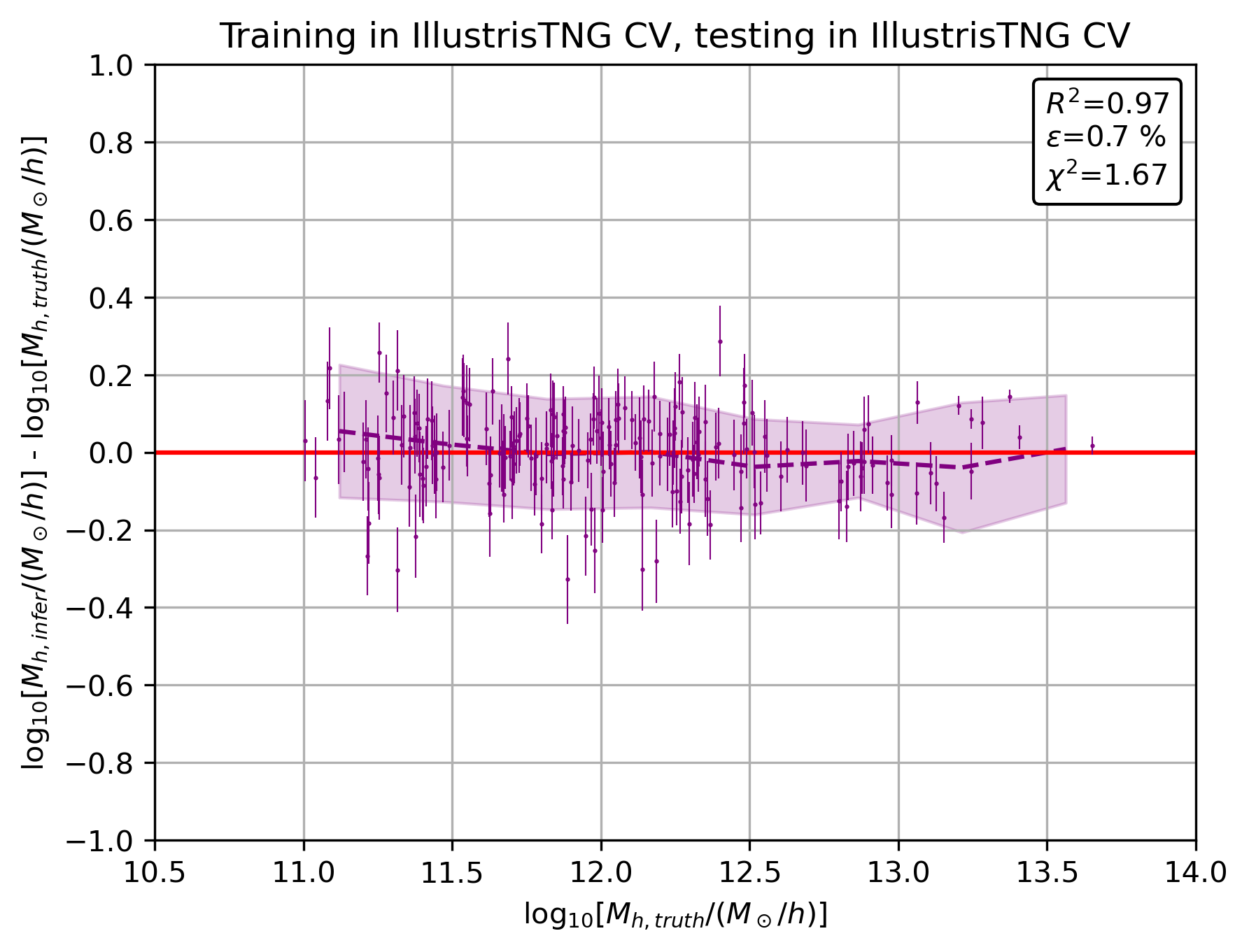}
\includegraphics[width=0.49\linewidth]{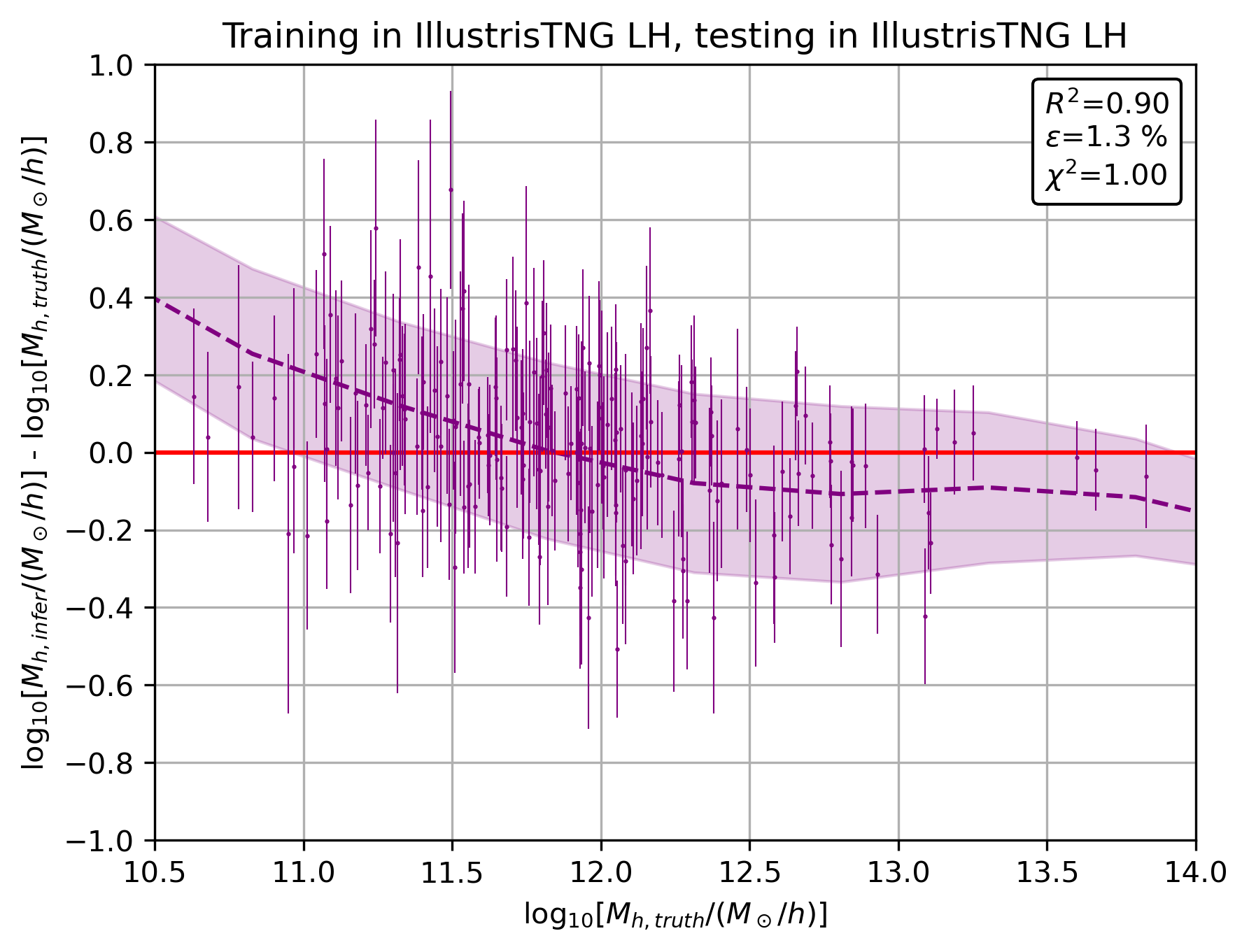}
\includegraphics[width=0.49\linewidth]{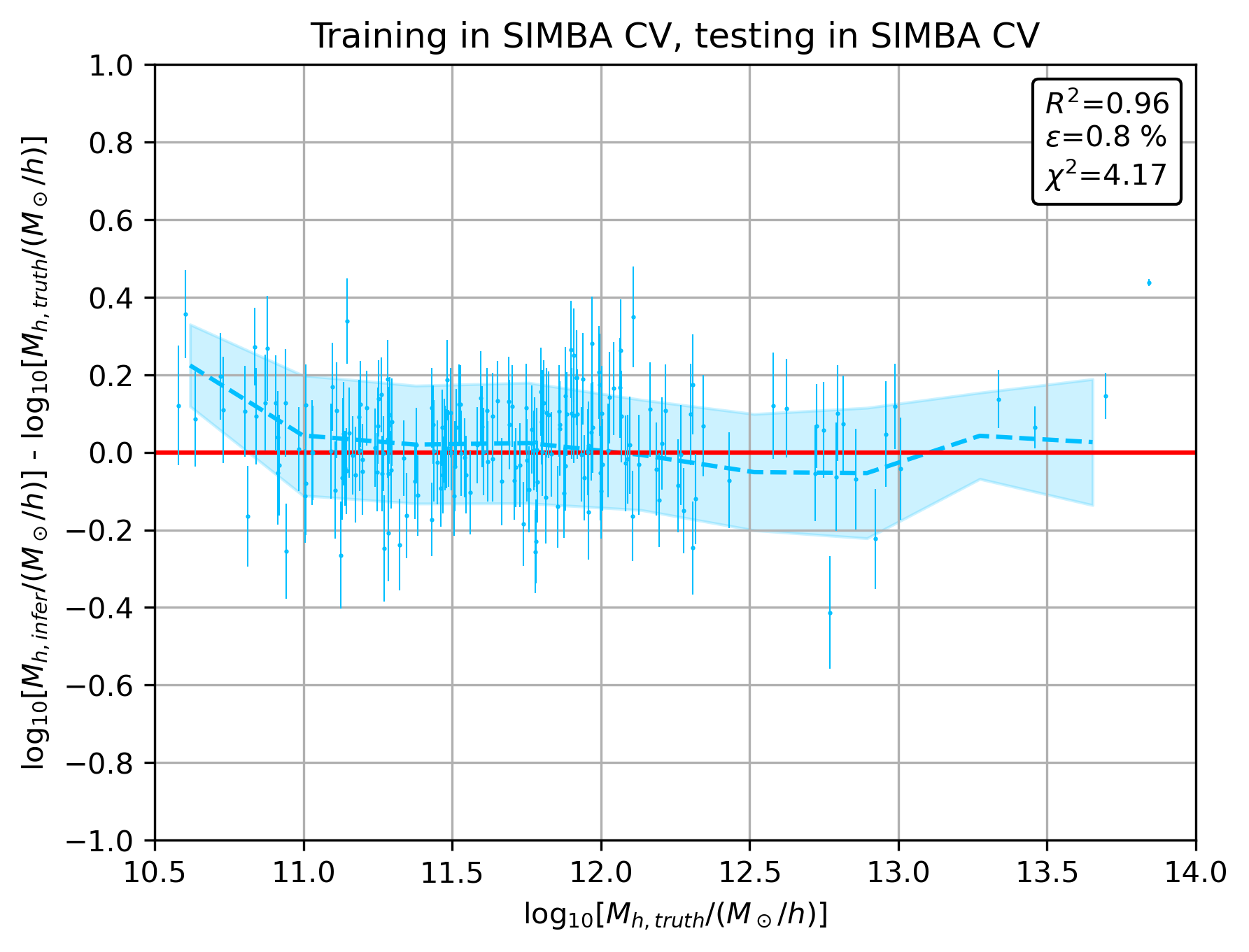}
\includegraphics[width=0.49\linewidth]{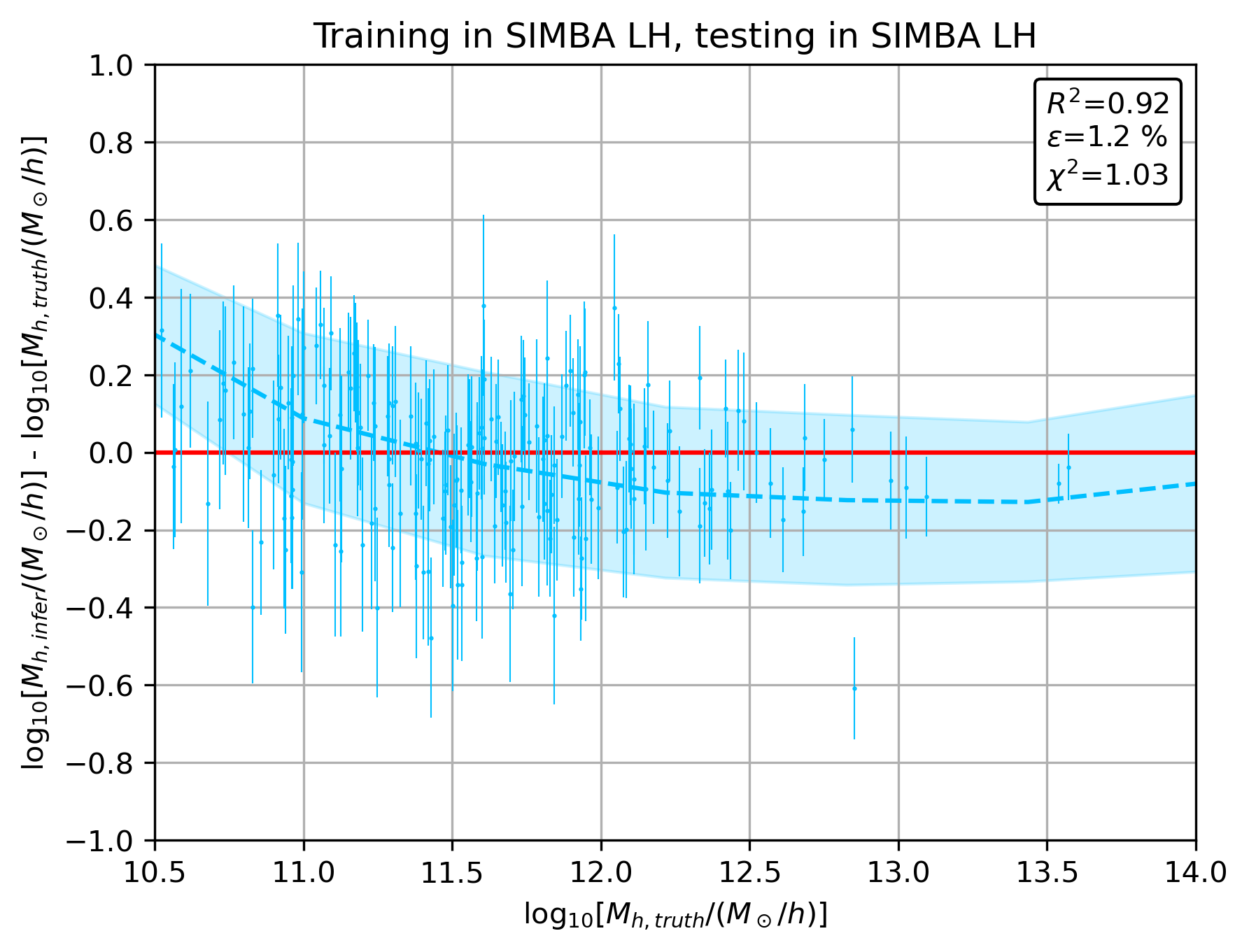}
\caption{Predicted versus true logarithm of halo masses $log_{10}\left[M_h/(M_\odot/h)\right]$ for the CV (left) and LH sets (right), training in the IllustrisTNG suite (top) and in SIMBA (bottom). A sample of 200 halos in the test dataset is shown in each case. Shaded regions and dashed lines correspond to real standard deviation and mean of test points, respectively. While in the CV set, astrophysical and cosmological parameters are fixed to fiducial values, the LH set comprises a broad range of astrophysical and cosmological scenarios. Even so, the GNN is still able to learn the halo/galaxy relation and predict masses in the LH case, only slightly worsening the prediction with respect to the CV case.}
\label{fig:out_true_illustris}
\end{center}
\end{figure*}

Here we first discuss the results of training a GNN to predict halo masses using galaxies from simulations of the CV set. As stated in Sec. \ref{sec:camels}, this set contains 27 simulations with fixed fiducial values for the cosmological and astrophysical parameters, only varying the random seed. The left panels of Fig. \ref{fig:out_true_illustris} show the accuracy of the network in the CAMELS CV sets, where the top panel stands for the IllustrisTNG subgrid model and the bottom one for the SIMBA suite. The vertical axis is the difference between the predicted and true logarithms of the halo mass, $\log_{10}\left(M_h/(M_\odot/h) \right)$, with respect to the true value in the horizontal axis. Error bars have been estimated via likelihood-free inference, sampling the standard deviation of the posterior, as outlined in Sec. \ref{sec:training}. One can see that the performance is fairly good in both cases, as the linear correlation coefficient of $R^2=0.96-0.97$ confirms, with a relative error lower than $\sim 1\%$ in $y$ ($\sim 25-40 \%$ in the mass itself). The dashed lines show the mean of test points, while the shaded region depicts their actual standard deviation, which extends up to $\sim 0.14$ dex. Thus, most of the test predictions lie within this region, although there are some outliers. The neural network is thus able to accurately infer the halo mass given some features of its galaxies.

Nevertheless, the previous results only show that the GNN is capable of predicting the mass of a halo when cosmology and astrophysics is known, i.e., when the parameters are fixed. However, the specific values of the relevant parameters for the real Universe are not well known yet, specially the astrophysical ones. In order to marginalize over uncertainties in cosmology and astrophysics, we have trained our network in the LH simulation set, which includes one thousand simulations varying cosmological and astrophysical parameters, together with the random seed (to also incorporate effects of sample variance), as noted in Sec. \ref{sec:camels}. The right panels of Fig. \ref{fig:out_true_illustris} show the GNN predictions for training the GNN in IllustrisTNG (top) and SIMBA (bottom). One can see that the results only slightly worsen from the equivalent case in the CV simulation set, with $R^2=0.90-0.92$ and relative errors of $\sim 1\%$. Uncertainties are also larger than in the CV set, roughly by a factor of 2, meaning that the model trained in the LH set offers lower precision. This worsening is expected since the network needs to marginalize over the astrophysical and cosmological parameters, differently from the network trained on the CV set. It should be noted that a small bias appears at low masses, below $\lesssim 10^{11} M_\odot/h$, which deviates up to 0.4 dex (also present in the SIMBA CV case). This could be explained due to the smaller number of low-mass halos in the datasets, preventing the network to properly learn in that range. Moreover, the expected fewer satellites in low-mass halos could also affect the predictions, since the GNN counts on less satellites to extract information. In any case, these results indicate that our model is able to learn the mapping between astrophysics/cosmology and the halo mass, and thus marginalize over the value of the parameters present in the simulations to make an accurate prediction.

To further evaluate the predictive power of GNNs, it is worth comparing these results to those obtained from the naive fit based only on stellar mass outlined in Sec. \ref{sec:benchmark}. For both the CV and LH sets, the GNN predictions outperform those from the polynomial fit, presenting larger $R^2$ coefficients and lower relative errors. The improvement is especially clear in the LH case, where the $R^2$ is significantly better than  the benchmark method. Moreover, the scatter around the true values spans $\sim 0.14$ and $\lesssim 0.2$ dex, compared to the larger standard deviations from the naive fit, which are $0.2$ and $0.3-0.4$ dex respectively. Note that $0.2$ dex is a factor up to $\sim 1.6$ in the mass (rather than in the logarithm), while $0.3$ dex is a factor $\sim 2$, meaning a $\sim 100$ \% error in the mass. These results demonstrate how taking advantage of the graph structure and further galaxy features, it is possible to attain richer correlations and better results.

Nevertheless, it has to be emphasized that the linear correlation coefficient $R^2$ and the relative error cannot constitute a complete statistical summary for testing the accuracy of the GNN. It is because our network is also predicting the standard deviation of the target, for which an additional component to the loss has been included, as discussed in Sec. \ref{sec:training}. Thus, neither $R^2$ nor the relative error quantify the sampling accuracy of the standard deviation. To test whether the uncertainties are reasonably predicted, it is useful to compute the $\chi^2$, defined as
\begin{equation}
    \chi^2 = \frac{1}{N} \sum_i^N \frac{(y_{{\rm truth}, i} - y_{{\rm infer}, i})^2}{\sigma_i^2}~.
\end{equation}
Note that minimizing the loss function contribution from Eq. \ref{eq:loss2} tends to drive the $\chi^2$ towards unity. This is actually the case in the LH set, where $\chi^2 = 1.00$ for IllustrisTNG and $\chi^2 = 1.03$ for SIMBA, indicating that uncertainties are accurately predicted. In the CV cases, however, while the mean predictions are more accurate, some errors are underestimated, leading to larger $\chi^2$ values, around $\sim 4$ in SIMBA. Moreover, since the test dataset is smaller, a few outliers with too small standard deviations greatly impact the value of the $\chi^2$.

One can also compute how many points in the test dataset present an accurate uncertainty by counting the fraction which fulfill the conditions $|y_{{\rm truth}, i} - y_{{\rm infer}, i}| \leq \sigma_i$ and $|y_{{\rm truth}, i} - y_{{\rm infer}, i}| \leq 2\sigma_i$, i.e., how many points lie within one and two times the standard deviation of the posterior. We find that for the LH cases, the fraction of points fulfilling the above conditions is 69\% and 95\% respectively for both suites. For the CV sets, these fractions only slightly deviate, 62 and 90\% for IllustrisTNG, and 72 and 96\% for SIMBA, respectively. For Gaussian distributed errors, these fractions should be around 68 and 95\% respectively. Note however that our calculation of the posterior mean and standard deviation does not make any assumption about the form of the posterior. Therefore, the numbers quoted above should be taken with caution and comparison with the Gaussian case should be done in a careful manner.

\begin{figure*}[th!]
\begin{center}
\includegraphics[width=0.49\linewidth]{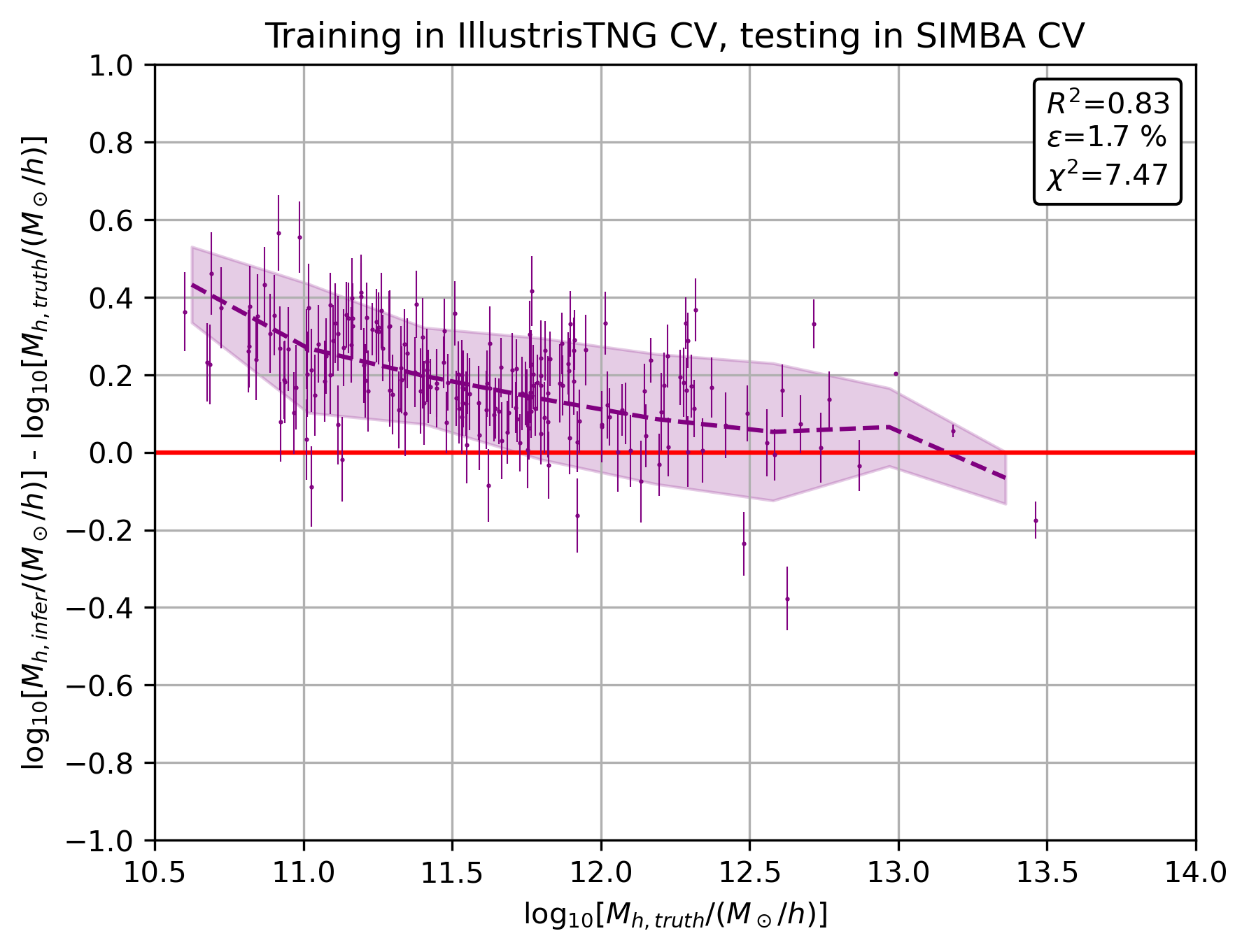}
\includegraphics[width=0.49\linewidth]{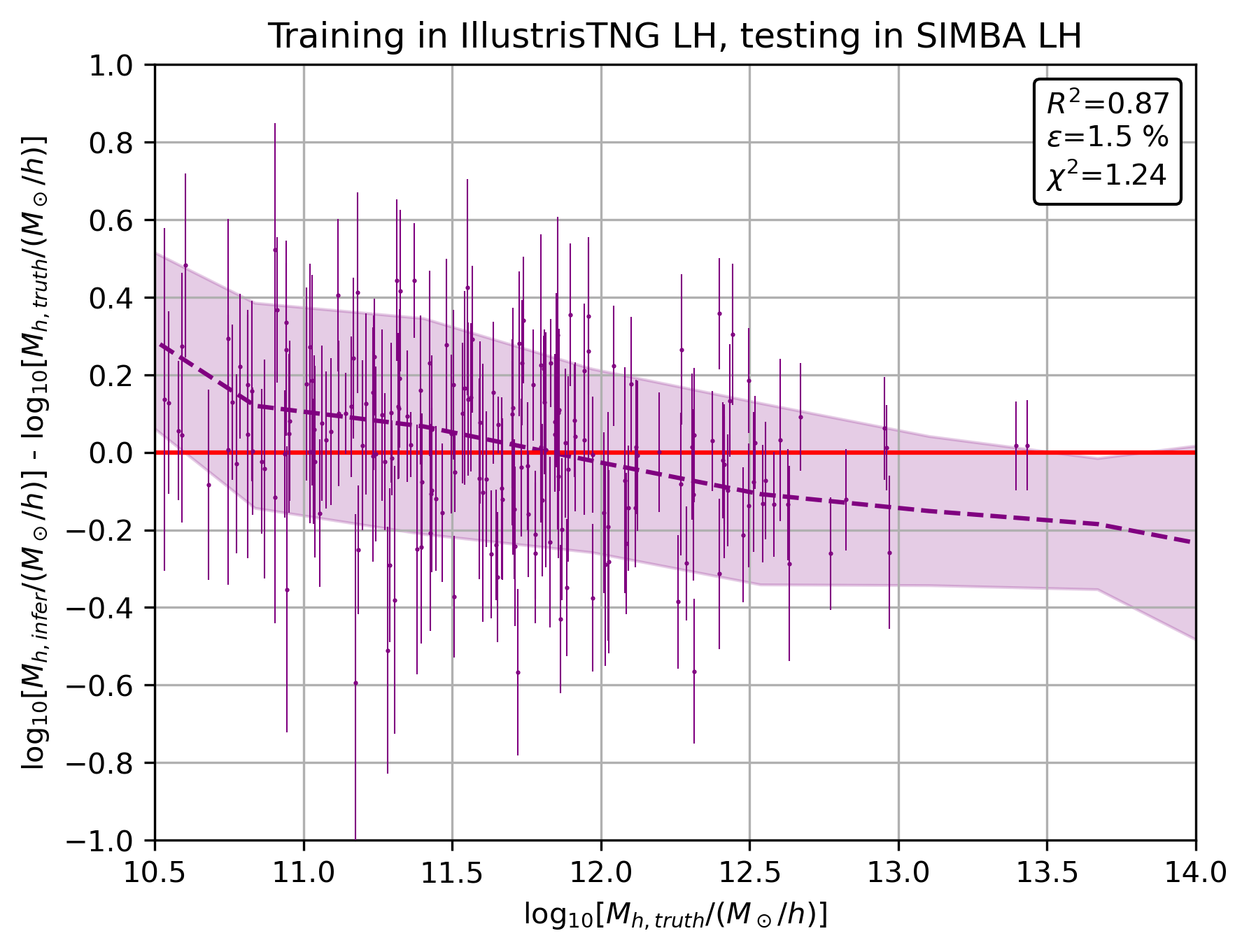}
\includegraphics[width=0.49\linewidth]{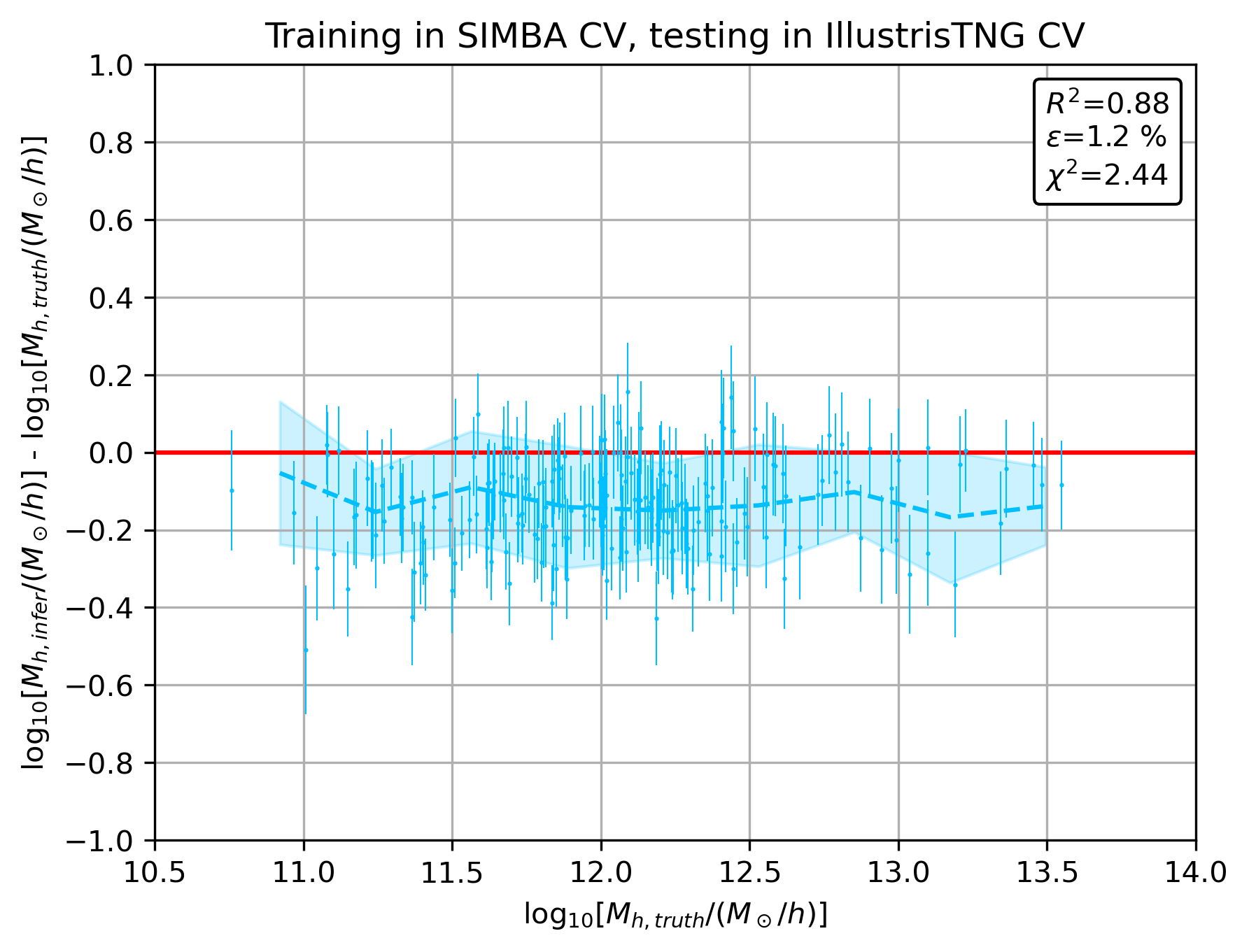}
\includegraphics[width=0.49\linewidth]{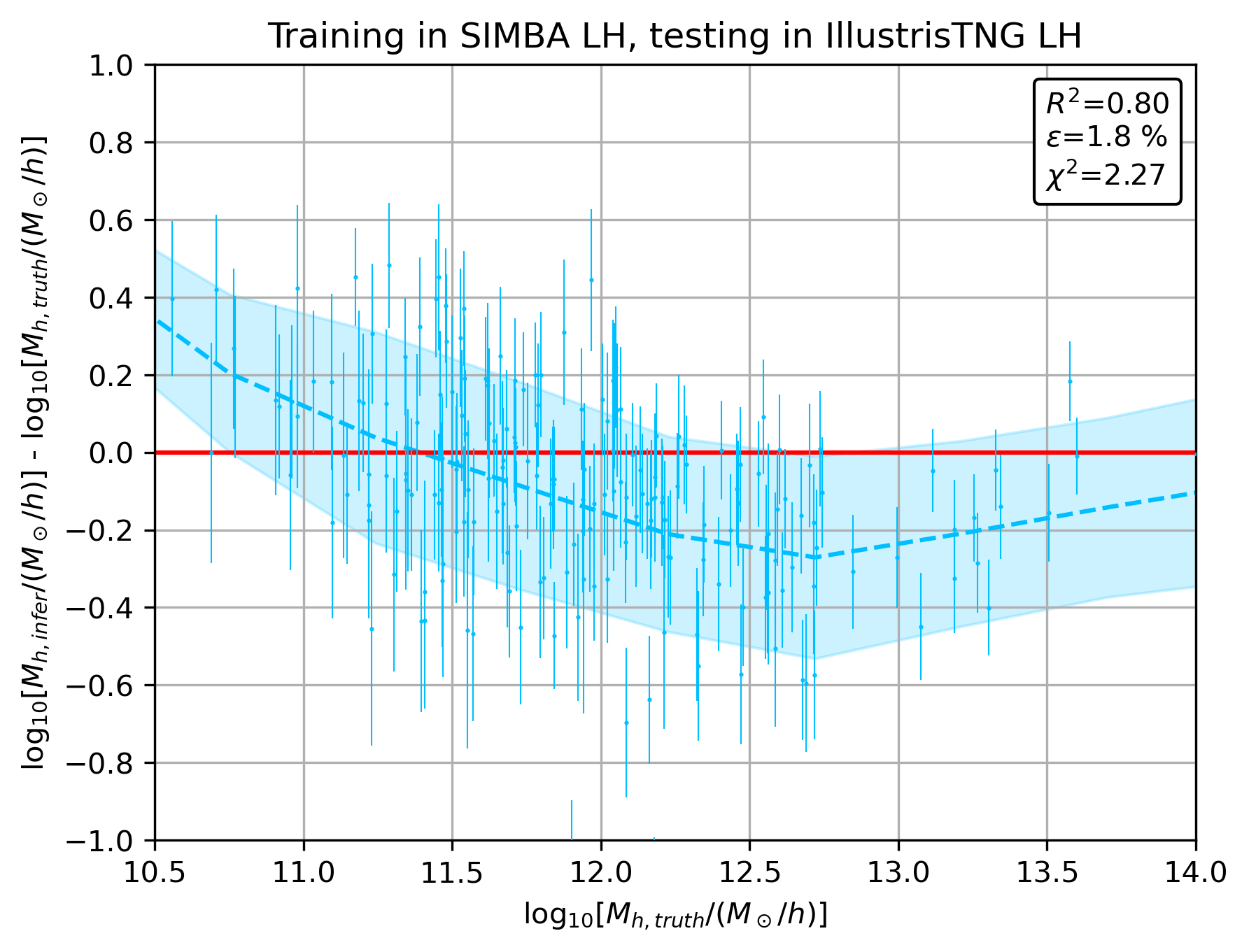}
\caption{Same as Fig. \ref{fig:out_true_illustris} but either using a model trained with the IllustrisTNG suite and tested with the SIMBA simulations (top) or trained in SIMBA and tested in the IllustrisTNG suite (bottom), for CV (left) and LH (right) sets. A model trained in a given suite worsens its behavior when tested in the other one, appearing biased in the CV case. However, in the LH set, it is possible to find a mapping between the parameter space of both subgrid physics models, alleviating such biases.}
\label{fig:out_true_simba}
\end{center}
\end{figure*}

There is another way to figure out whether uncertainties are correctly sampled. The shaded regions in Fig. \ref{fig:out_true_illustris} represent the actual standard deviation of the test points computed within several mass bins. Therefore, if the predicted uncertainties are accurately sampled, their mean value $\overline{\sigma}$ should correspond to those shaded regions. For the CV case, despite the GNN providing more accurate models, mean uncertainties $\overline{\sigma}$ are underpredicted with respect to actual scatter by $\sim 40$ \% for IllustrisTNG and $\sim 20$ \%  for SIMBA. However, in the LH case, uncertainties are better sampled, only deviating $\sim 3$ \% and $\sim 7$ \% for IllustrisTNG and SIMBA respectively. This implies that models trained in LH, despite predicting slightly less accurate results than those from the CV set, provide a better sample of the posterior uncertainties.

\subsection{Robustness over different subgrid physics models}
\label{sec:robust}

Subgrid physics, i.e. the models used to simulate unresolved astrophysical processes such as the feedback from supernovae and black holes, can only be implemented in a phenomenological way and therefore there is not a unique subgrid model that best represents reality. Thus, having a ML model robust over different subgrid scenarios would be needed in order to obtain predictions that do not depend on the particular type of simulation used to train the networks. 

To check whether our GNN fulfills this requirement, we have taken the model previously trained on simulations from the IllustrisTNG suite, and we have tested it on galaxies from the SIMBA simulations, which make use of a completely different subgrid physics model for AGN and SN feedback. The top left panel of Fig. \ref{fig:out_true_simba} shows the predictions for the halo mass in the CV set, i.e., trained in IllustrisTNG CV and tested in SIMBA CV. We see that the performance becomes worse, and actually a bias appears. This offset may arise from the fact that the IllustrisTNG and SIMBA models make different predictions for the halo-galaxy connection, as shown in Fig. \ref{fig:scatplot}. This can be related with the fact that astrophysical parameters are not completely correlated and calibrated between both cases. While the CV set assumes fiducial values for the astrophysical and cosmological parameters, the default values in the IllustrisTNG suite do not correspond to the ones in SIMBA, since they refer to different quantities and physics. The absence of a one-to-one relation between both suites in the CV set may explain why the network fails to make a robust prediction.

The right panel of Fig. \ref{fig:out_true_simba} depicts the results of testing the network trained on galaxies of the IllustrisTNG LH set on galaxies of the SIMBA LH set. In this case, the bias present in the CV case disappears, and only a broad scatter holds. The absence of the offset can be attributed to the intrinsic marginalization of astrophysical effects carried out by the network (although a slight tilt is present in the trend). Moreover, the linear correlation coefficient only slightly decreases with respect to testing in IllustrisTNG (top right panel of Fig. \ref{fig:out_true_illustris}), and is actually better than in the CV counterpart (top left panel of Fig. \ref{fig:out_true_simba}), in spite of dealing with a much broader astrophysical parameter space. The $\chi^2$ also remains closer to unity than in the CV case, meaning that uncertainties are better sampled in the LH set. The fraction of points fulfilling the conditions $|y_{{\rm truth}, i} - y_{{\rm infer}, i}| \leq \sigma_i$ and $|y_{{\rm truth}, i} - y_{{\rm infer}, i}| \leq 2\sigma_i$ is 65 \% and 93 \% respectively, while much lower for the CV case (20 and 46\% respectively). These facts imply that models trained in the LH set generalize better than those trained in the CV one. 

An analogous experiment has been carried out, employing the model trained in the SIMBA suite and testing it in IllustrisTNG. The results are shown in the bottom panels of Fig. \ref{fig:out_true_simba} for the CV set (left) and LH set (right). Note the offset in the CV case, which is similar but opposite (underpredicting the mass) to the one appearing in the top left panel of Fig. \ref{fig:out_true_simba}. This is reasonable, since given that the IllustrisTNG model overpredicts the mass in SIMBA, the contrary should be expected when a SIMBA model is tested in IllustrisTNG. In the LH scenario, predictions are slightly worse, decreasing the truth-prediction correlation down to $R^2=0.8$, and also poorer than in the CV case. Both sets present larger $\chi^2$ values and a deviation from the Gaussian counterpart, given that only a fraction of $\sim 50$\% and $\sim 80$ \% points lie within one and two times the posterior standard deviation, respectively.

It is noteworthy that, for the LH set, a model trained in IllustrisTNG generalizes better in SIMBA than the opposite case. An interpretation of this outcome is problematic. SIMBA usually covers more extreme astrophysical scenarios than IllustrisTNG, presenting also more galaxies per halo (see Fig. \ref{fig:histogram}), facts which could have an impact in this different cross testing. SIMBA simulations usually show a more stochastic behavior, presenting more scatter in some properties such as galaxy scaling relations. This can be due to the small box size compared to the large scale effect of the SIMBA AGN jets \citep{Dave:2019yyq, villaescusanavarro2020camels}. Nevertheless, although we observe a higher variance and some outliers, most of the predictions only deviate up to 0.4 dex from the ground truth. In all cases, the mean relative error in the logarithm of the mass lies below 2\%. The GNNs are able to recover the true mass in the majority of the cases within the standard deviation uncertainty. This shows that the models are relatively accurate even when they are applied to simulations with a different subgrid physics modeling, manifesting their robustness.

\begin{figure}[t!]
\begin{center}
\includegraphics[width=0.99\linewidth]{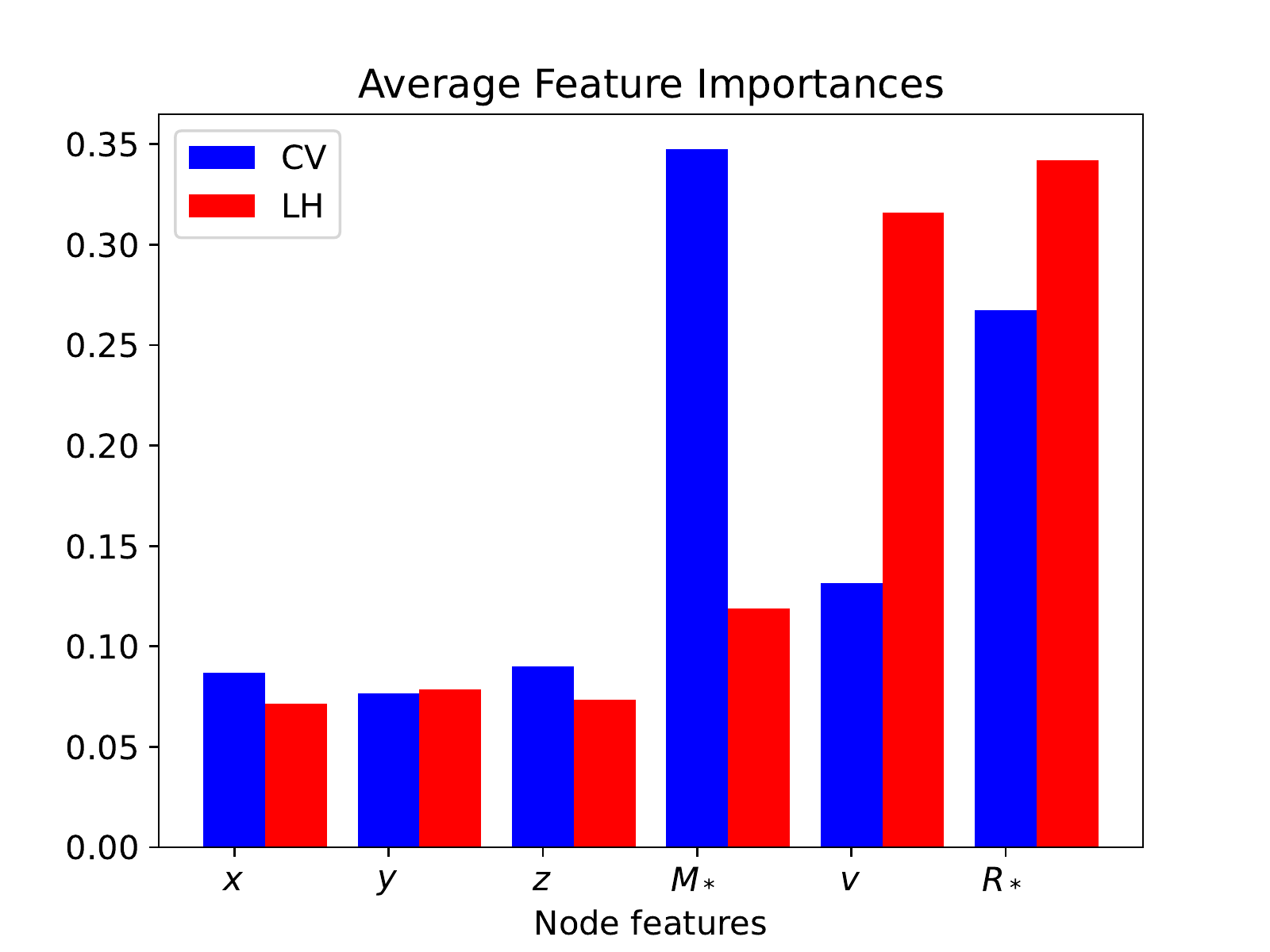}
\caption{Saliency values of the galaxy features employed to train the GNN. Larger values imply that the network prediction
is more influenced by variations of that feature, indicating that such properties are more relevant for the result.}
\label{fig:captum}
\end{center}
\end{figure}

\section{Interpretability of the GNN}
\label{sec:interpret}

\begin{figure*}[th!]
\begin{center}
\includegraphics[width=0.32\linewidth]{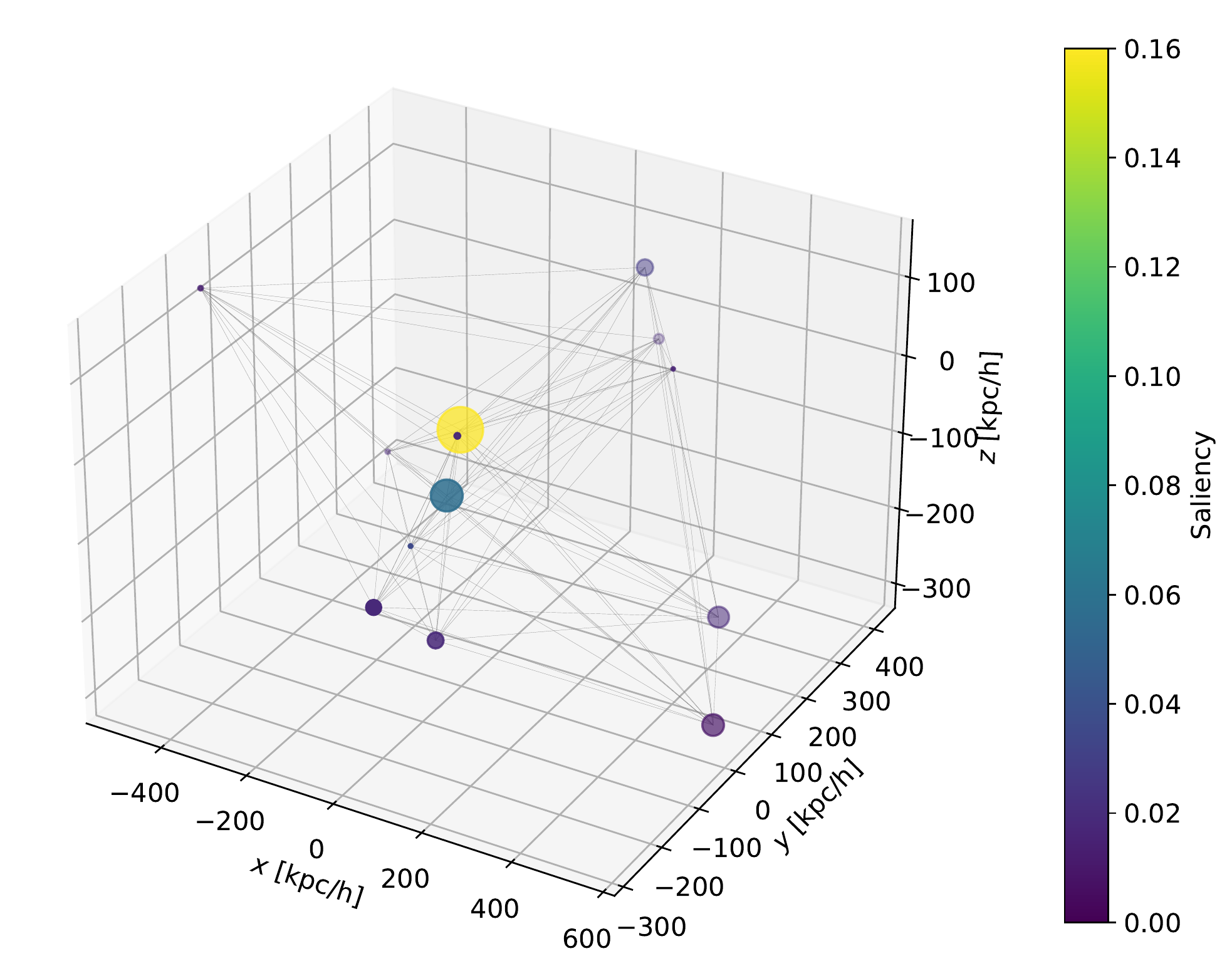}
\includegraphics[width=0.32\linewidth]{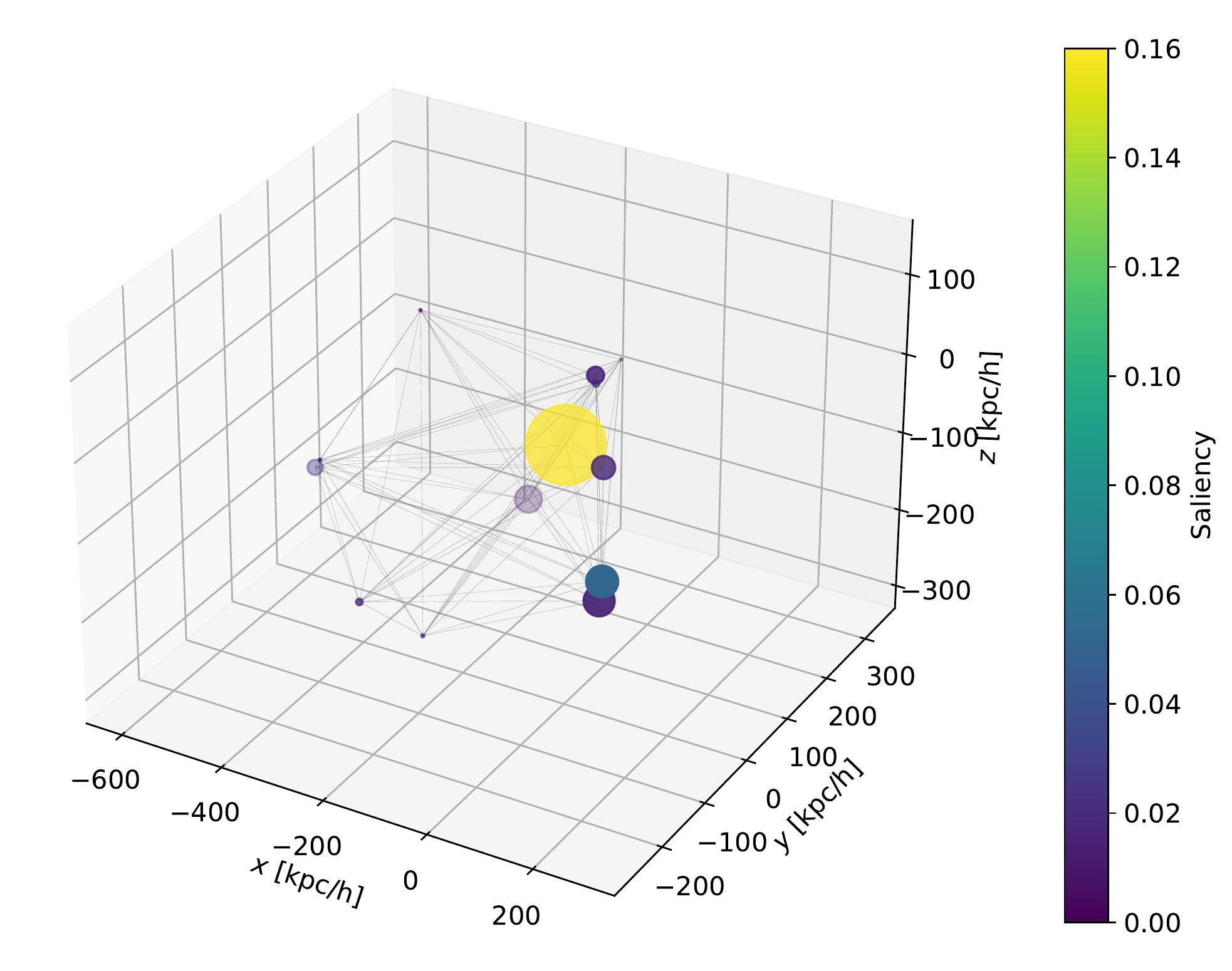}
\includegraphics[width=0.32\linewidth]{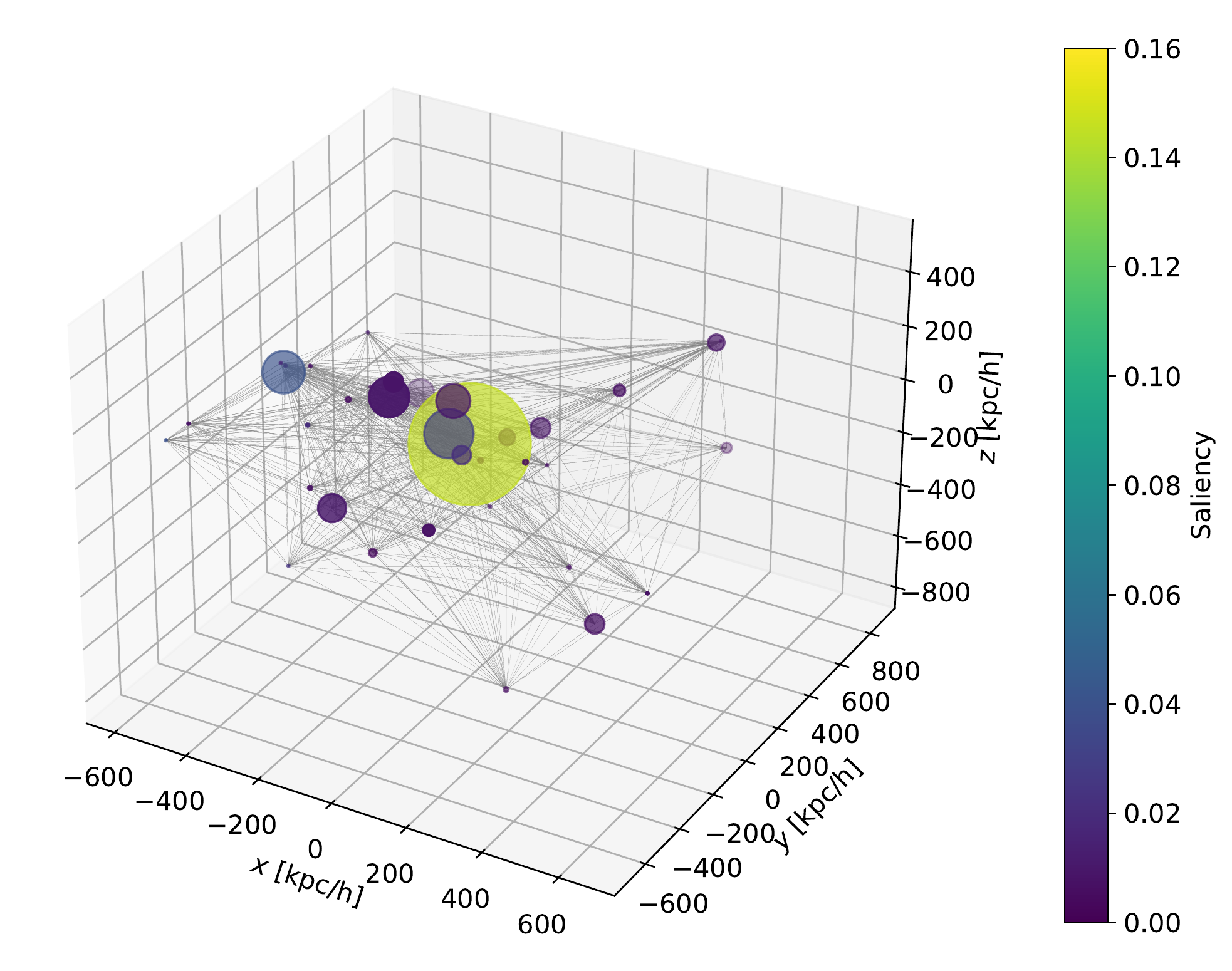}
\includegraphics[width=0.32\linewidth]{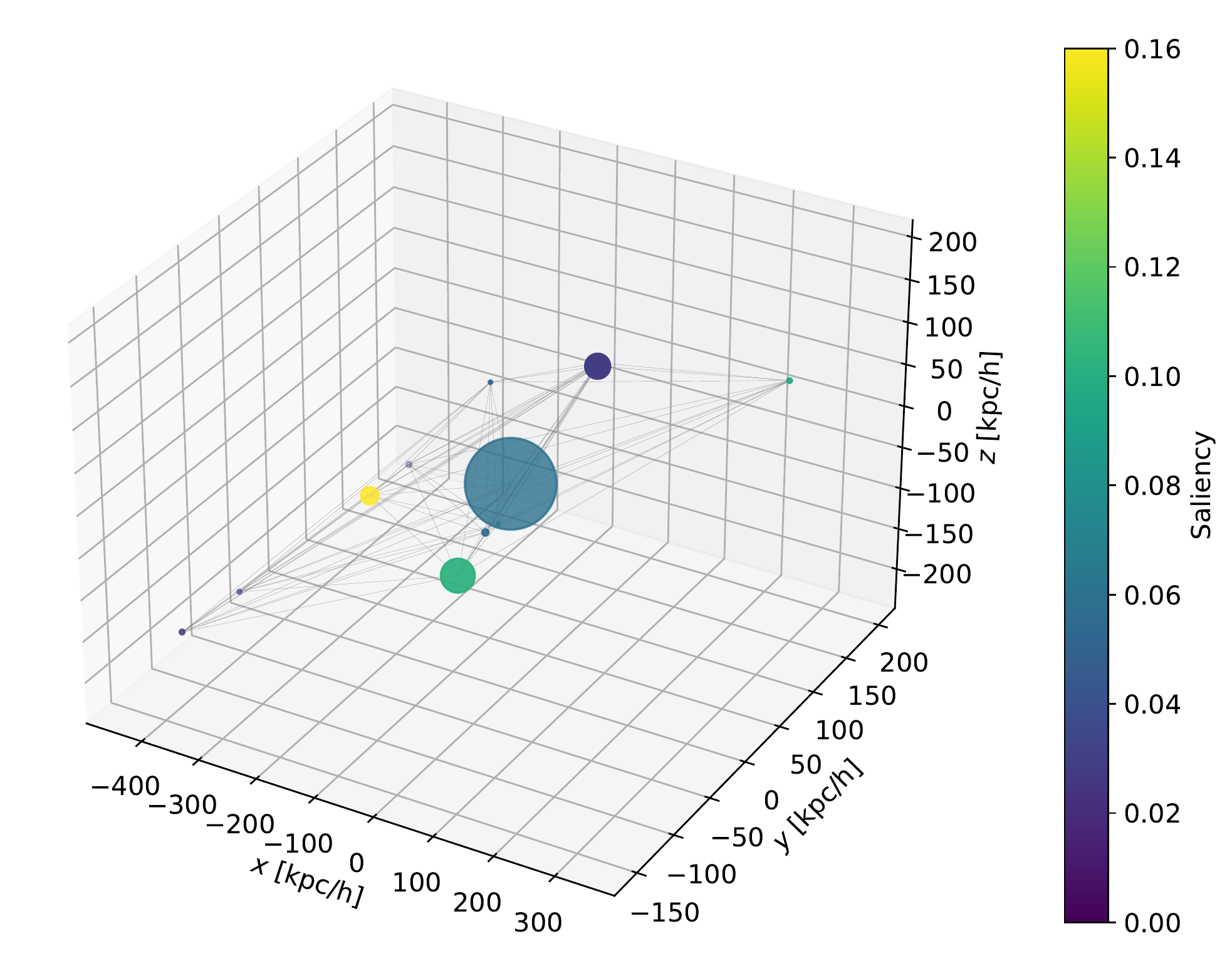}
\includegraphics[width=0.32\linewidth]{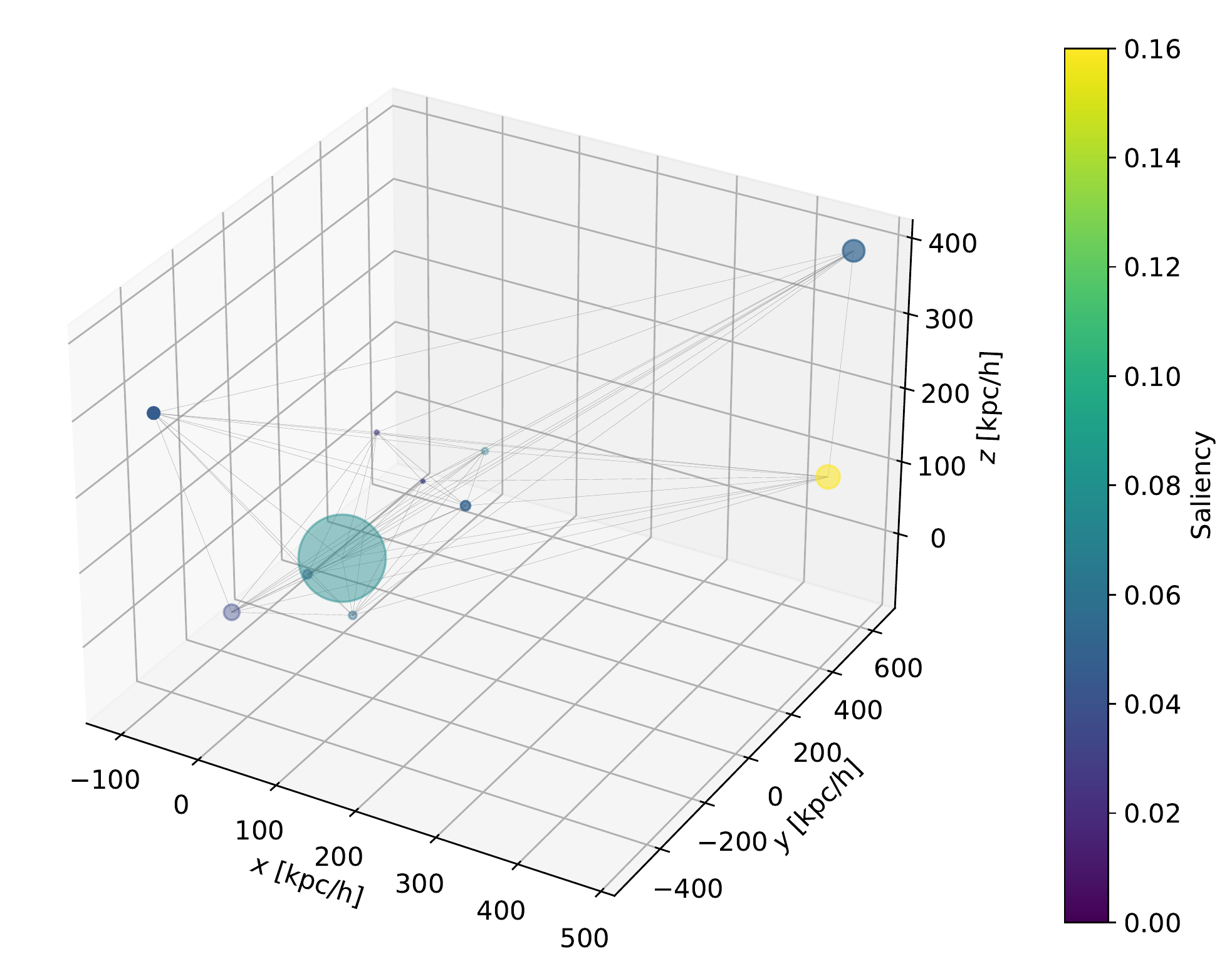}
\includegraphics[width=0.32\linewidth]{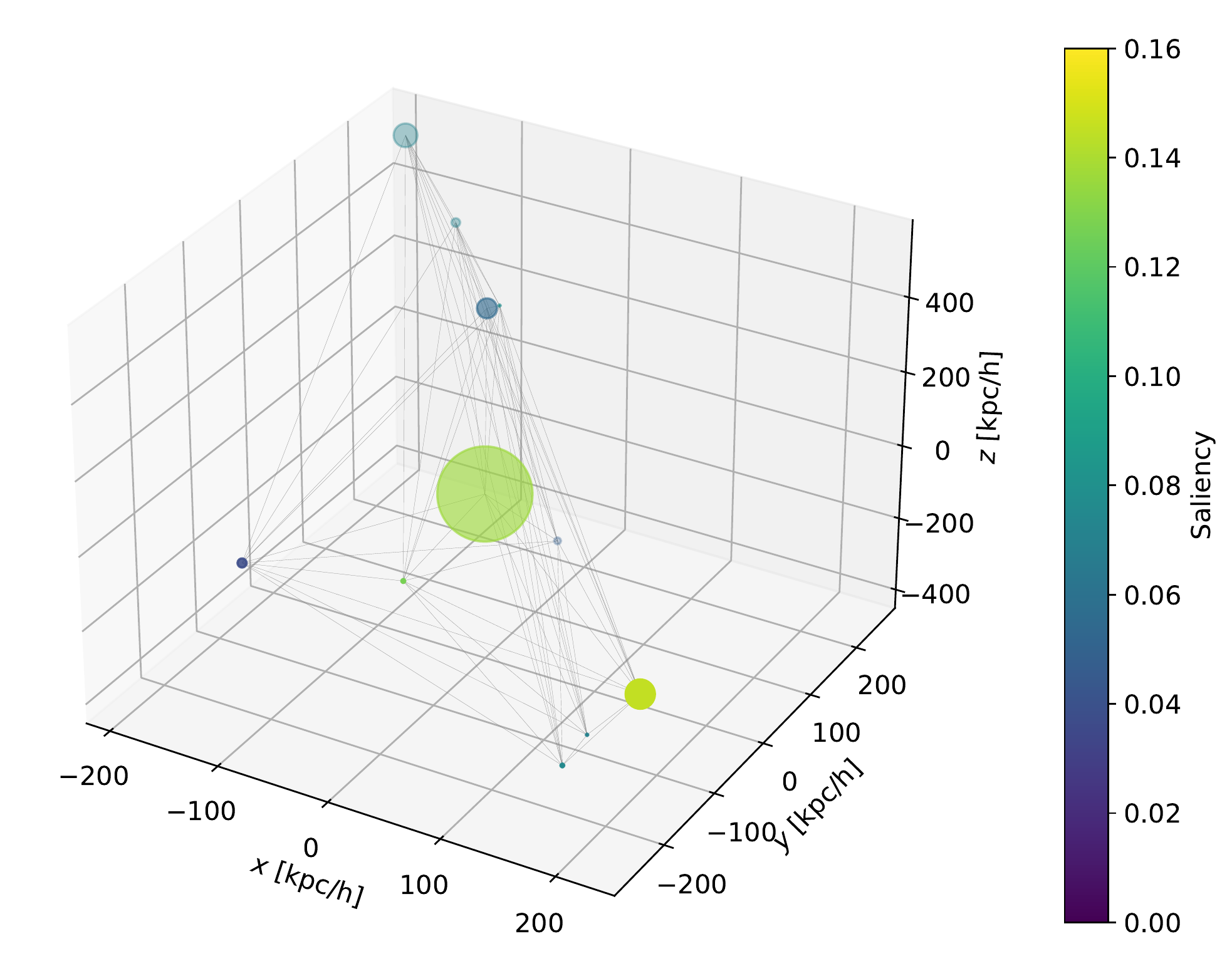}
\caption{Saliency graphs for six different halos from the IllustrisTNG CV (top) and LH (bottom) datasets. Colors denote the saliency attributes, meaning that galaxies with larger values are more relevant for the prediction. Sizes of the nodes are proportional to the stellar mass of the galaxies. In the CV set, central galaxies usually have the greatest impact on the prediction, although if there are massive satellites, they can also contribute significantly (e.g., right panel). In the LH case, however, more attention is often focused on smaller satellites rather than on central galaxies.}
\label{fig:saliencygraphs}
\end{center}
\end{figure*}

Neural networks represent superb tools to deal with multiple problems, but the factors that determine their behavior and performance are difficult to understand. It is always desirable to gain some interpretability of our ML models, in order to determine which features and inputs are the most relevant for predicting the output.  To do so, we make use of the Python library Captum\footnote{\url{https://captum.ai/}} \citep{kokhlikyan2020captum}, a package designed for model interpretability and attribute selection. Specifically, we employ the saliency method for computing the gradients of the outputs with respect to the inputs and features employed \citep{2013arXiv1312.6034S}. In that way, larger gradients, and hence larger saliency values, imply that the prediction is more sensitive to variations of that variable.

We have computed the gradients with respect to the features employed in the training of the GNN, averaged over all galaxies and taken the absolute value. This approach gives us a saliency value for each property, which is shown in Fig. \ref{fig:captum} for the IllustrisTNG suite in the CV (blue) and LH (red) sets. For both sets, we can notice that, as could be expected, each coordinate of the position presents a very similar importance, due to implicit rotational invariance present in the problem. Regarding the CV case, the most relevant features to determine the output appear to be the stellar mass, $M_*$ (as expected from the tight correlation shown in left panel of Fig. \ref{fig:scatplot}) and subsequently, the stellar half-mass radius, $R_*$. However, in the LH case, the network focuses more on $R_*$ and $v$ rather than the stellar mass. One can interpret this change of learning behavior from the larger scatter in stellar mass shown in Fig. \ref{fig:scatplot} in LH with respect to the CV set. In other words, the large variety of astrophysical and cosmological scenarios in the LH set may require the network to base its predictions on features that do not exhibit such large scatter. While the predictor mostly rely on the $M_*$-$M_h$ correlation when the scatter is small, velocities and galaxy size become better tracers otherwise. Nevertheless, one has to be cautious with these interpretations, since the saliency values may not give a complete picture. For instance, spatial position importance may be underestimated since its weight could be split into the three coordinates.

It is common in computer vision tasks to calculate the saliency map, which indicates those pixels in an image that are most relevant for the final output (see, e.g., \citealt{2021ApJ...907...44V} for an application in cosmology). In our case, we are dealing with graphs rather than images. Therefore, for a given halo it is possible to compute the \textit{saliency graph}, which shows the nodes whose features are more relevant for predicting the halo mass. Saliency values can be computed using the same procedure as above, but taking the absolute value and averaging over all features at each node. Examples of such saliency graphs for different halos are depicted in Fig. \ref{fig:saliencygraphs}, where the color indicates the saliency value, and the size is proportional to the stellar mass. Neighbors are connected by lines. Chosen samples present relative errors lower than 1.5 \% to ensure that their saliency graphs are meaningful. Top row stands for the CV set and bottom for LH, both in IllustrisTNG. In the CV set, as one would naively expect, the central galaxies provide the most relevant nodes for the output. These galaxies are also those with larger stellar masses. However, given that the stellar mass is an informative property, as seen in Fig. \ref{fig:captum}, halos with relatively massive satellites can also show other relevant nodes besides the central one. On the other hand, since in the LH set, stellar mass is less informative, as seen in Fig. \ref{fig:captum}, the network may focus more on some low mass satellites rather than in central galaxies, which become less important. This illustrates the importance of the satellite population in our method. Hence, one has to be cautious when applying these models to galactic systems, since excluding some relevant satellites may have an impact in the predictions and induce a bias in the results.

\begin{figure}[t!]
\begin{center}
\includegraphics[width=0.99\linewidth]{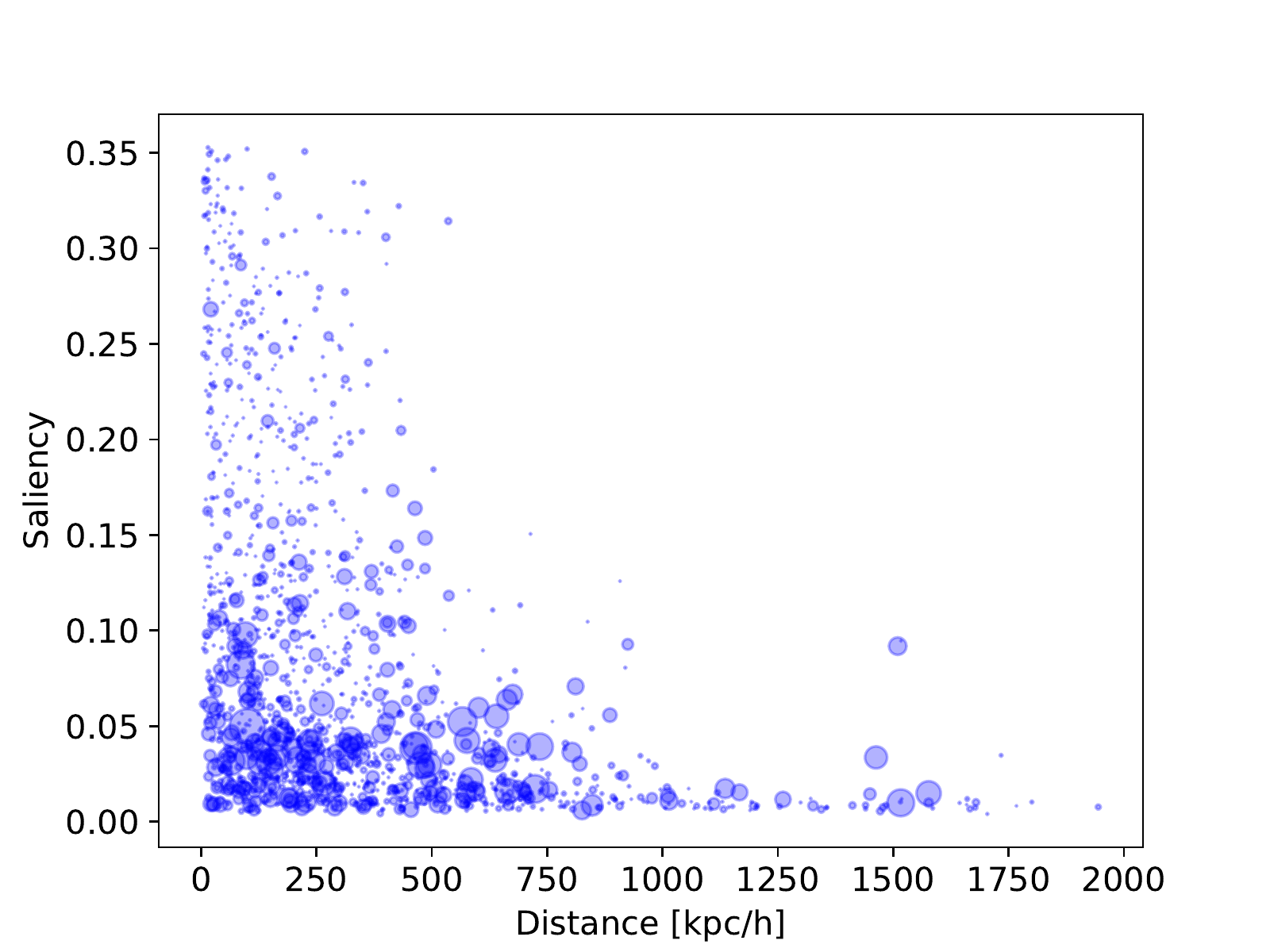}
\caption{Saliency for each satellite galaxy depending on its distance to the halo center, excluding the central ones. Point diameter indicates the stellar mass of the nodes. There is a general tendency where closer galaxies to the center present larger saliency values, and therefore, are more relevant for predicting the output.}
\label{fig:distances}
\end{center}
\end{figure}

It is thus pertinent to ask ourselves which satellites leave a greater impact on the halo mass. Fig. \ref{fig:distances} depicts the saliency value of each satellite galaxy as a function of distance to the center, excluding central galaxies. Point size is proportional to the stellar mass of the node. One can notice a trend where closer galaxies become more relevant than those farther away. Moreover, this tendency seems to be relatively independent on their stellar mass. Furthermore, in real applications, the membership of galaxies lying at the boundaries of the system may be less secure, being not clear whether a far galaxy is a satellite or not. Fig. \ref{fig:distances} shows that removing or including such candidates would not have a significant impact in the results of the network, given the decreasing saliency value with the distance. This plot employs the IllustrisTNG CV dataset, although similar qualitative conclusions can be extracted from the other cases.

These tests provide us with enlightening information about how GNNs learn and predict their outputs, as well as which are the most relevant components to understand the halo/galaxy connection. It has to be noted that the previous interpretations have to be taken with caution, since the saliency maps can be sometimes dominated by noise in the gradients and lose meaningfulness. Further development is required in order to obtain a better understanding of the training procedure and the emergent properties of halos from galaxies.

\section{Discussion}

\label{sec:discussion}

\subsection{Main conclusions}

Constraining the total mass of dark matter halos from galaxies still presents a challenge from both theoretical and observational perspectives, given the large contribution of the dark matter component. In this work we have presented a new method based on artificial intelligence designed to infer the total mass of a halo from the properties of the galaxies it hosts. The point cloud arrangement of halo-galaxy catalogues has been exploited in order to structure halos as mathematical graphs, where galaxies constitute the nodes, connected by proximity. This organization of data makes it possible to employ GNNs, naturally suited to operate with graphs, to extract global permutation invariant quantities, as is the case of the halo mass. The models are fed with different observable galactic features, such as the position, velocity, stellar mass, and stellar half-mass radius.

We have made use of the large collection of state-of-the-art hydrodynamic simulations from the CAMELS project to train our networks, which include thousands of simulations covering different astrophysical scenarios. Training GNNs over this dataset allows us to achieve precise models capable of predicting the mass of a halo with remarkable accuracy. Here we outline some of the key conclusions of our method:

\begin{itemize}
    \item Our models are able to accurately infer the mass of a halo when trained and evaluated in simulations with fixed astrophysical and cosmological parameters, with a precision better than $\sim 1$\% (in terms of the logarithm of the halo mass).
    \item The networks have also been trained in simulations with different astrophysical and cosmological scenarios, successfully marginalizing over a broad astrophysical parameter space, and learning the connection between the halo mass and the properties of the galactic components. This provides accurate predictions of the total halo mass, with relative errors (in the logarithm of the halo mass) around $\sim 1$\%.
    \item We have proven that the trained networks in a simulation suite still provide relatively precise predictions when tested in simulations with a different subgrid physics model, only increasing the mean relative error up to $\sim 2$\%. This illustrates the robustness of this method with respect to the astrophysical modeling.
    \item Our results strongly rely on the velocity and size of satellite galaxies, demonstrating the importance of other galactic features beyond the stellar mass.
    \item We have performed likelihood-free bayesian inference, providing additionally an estimate for the standard deviation without knowing the actual likelihood.
    \item Hyperparameter optimization has been carried out to maximize the performance of the networks. 
\end{itemize}

\subsection{Comparison to previous works}

It is not straightforward to compare our results to previous works, given that the datasets employed, problem setups and features considered can be notably different. In the following, we compare our approach to other ML methods estimating the halo mass in previous literature, while emphasizing that a proper comparison between two techniques would require the use of common data and problem statement.

\begin{itemize}

\item \cite{2019MNRAS.490.2367C} apply several ML techniques, such as standard MLPs, to predict the halo mass from galaxy groups data. They train their models in semianalytical galaxy catalogues calibrated to SDSS data, outperforming traditional methods to infer the dynamical mass. Their $1\sigma$ scatter region spans $\sim 0.4-0.6$ dex (a fractional difference of $\sim 3 - 5$\%), while for the same masses, our GNN reduces down to $\sim 0.14$ dex for the CV set and $\sim 0.2$ dex for the LH one. In terms of the relative error in the mass (rather than in the logarithm), their scatter gets up to $\sim 250-400$\%, while our models reduce it down to $40$ \% and $60$ \% for CV and LH respectively. Furthermore, our models make use of fewer galaxy features (6 rather than their 9) and the LH simulations cover a large volume of the astrophysical and cosmological parameter space.

\item \cite{2019ApJ...881...74M} face a similar problem of predicting halo masses from group galaxy properties, splitting their datasets in red and blue groups, according to the color of the central galaxies. They train a random forest estimator, obtaining scatter comparable to ours (note the $\sqrt{2}$ factor of difference in their definitions), but employing 9 features with a semianalytical galaxy formation model with fixed parameters.

\item \cite{2021arXiv211101185V} predict galactic properties, such as the total mass, training different ML models in the TNG100 simulation. As in our case, the galaxies considered present stellar masses above $\gtrsim 10^8 M_\odot/h$, although they predict the subhalo mass rather than the halo one. They employ up to 15 features, including photometric, kinematic, and structural properties. When all features are included, they are able to get accurate results, with correlation of $R^2 =0.92$. This case can be compared to our IllustrisTNG CV set, since it presents fixed astrophysical and cosmological parameters,\footnote{Although TNG100 has a box size of $\sim$100 Mpc/$h$, larger than the 25 Mpc/$h$ CAMELS box size, the CV set includes 27 simulations varying the random seed.} where we obtain a better correlation of $R^2 =0.97$ even using only 6 features. Moreover, we are also capable of obtaining a similar accuracy in the LH set than their models in TNG100.

\item \cite{2020arXiv201110577L} train CNNs on the density field of dark matter halos to infer their masses, with an accuracy comparable to the one of our network train on the CV set. However, note that this is a very different task, since that method relies on the 3D dark matter density field rather than on observable features, as is the case of our GNNs. Moreover, the authors employ N-body simulations with fixed cosmology rather than full hydrodynamic simulations. In any case, one has to be cautious at comparing these different approaches, due to the distinct datasets, assumptions and features considered, being appropriate only for illustrating the potential of GNN models.

\end{itemize}

\subsection{Caveats and future prospects}

While the ML method presented here shows reasonable accuracy, it however has some caveats that have to be emphasized. For instance, it may not be obvious whether one galaxy belongs to a halo or not, in cases where two galactic groups are close together. This can be exacerbated in real observations, which take place in redshift-space. This may cause the appearance of interlopers in a halo which could be counted as its own satellites, distorting the results. The effect of the presence of other halos in the environment when building the graph is also disregarded, which could have an impact for some close halos gravitationally bound. A way to deal with the influence of surrounding groups may be to include some global feature regarding the amount of galaxies or halos within a certain distance from the halo which is evaluated.

Note that, since only halos with more than one galaxy have been considered, this method should only be applied to halos that contain multiple satellites (above the assumed mass threshold). In those cases where the existence of satellites is not clear, the method may not be reliable. Machine learning approaches could also be employed to identify the satellites around a given central galaxy, i.e., training a network to predict whether a subhalo is part of a given halo or not.

It is important to highlight that, given the small size of the simulation boxes, the amount of very massive galaxies may be undersampled. The fraction of galaxies of galaxies with stellar mass above $10^{11} M_\odot/h$ is $\lesssim 1\%$ for both suites, implying a relatively low amount of those large objects. Training in larger simulation boxes with a higher amount of large halos would be advisable to ensure the reliability of our method when applied to very massive galaxies.

Moreover, we have only trained with halos at $z=0$, but it would be also convenient to derive models capable of inferring halo masses at earlier times. New difficulties may arise in that case, such as redshift space distortions, or the fact that we could expect fainter galaxies at higher redshift (since for a given luminosity threshold, the number density of galaxies decreases with redshift), reducing thus the size of the training dataset.

The cross-tests discussed in Sec. \ref{sec:robust} are aimed to illustrate the robustness of the network under different conditions regarding the baryonic feedback modeling. While our results are still relatively accurate even when they are tested in a different subgrid physics model than the one used for training, it would be desirable to further maximize their robustness. While training on IllustrisTNG and SIMBA simultaneously could enhance the performance, there is no guarantee that the network would be more robust under other conditions. It is possible that the model learns some kind of bimodal distribution of simulations, identifying first whether the halo corresponds to a IllustrisTNG or a SIMBA simulation, and then predicting the halo mass. A further improvement on this would be to train on IllustrisTNG plus SIMBA and cross-test in a third different suite. Such additional suite with equivalent features to those of IllustrisTNG and SIMBA is still missing, although the CAMELS collaboration is already working to include other suites using different hydrodynamic codes and baryonic feedback models.

The results presented here are obtained by restricting ourselves to a reduced set of several observable quantities. However, it is possible in principle that using further variables, the results could improve. Additional correlations with the halo mass could arise if supplementary features are considered to train the GNNs. For instance, previous works like \cite{2021arXiv211101185V} have shown the importance of photometric variables, such as luminosities at different wavelenghts, for inferring the total halo mass. Additionally, considering, e.g., radii enclosing 20\% and 80\% of the mass (or light) may provide further information about the concentration and morphology of the halo. Among other appealing galactic properties that could be contemplated are the gas mass, the HI mass, metallicities, velocity dispersion, etc. Furthermore, it would be desirable to explore more complex GNN architectures to find whether more accurate results can be achieved, reducing the scatter and the predicted uncertainties. The training of GNNs on galactic properties combined with other observables such as lensing could also enhance its predictive power.

The framework developed in this article has been proposed to infer the total mass of a given halo. However, since it is based on extracting global quantities from galaxy features, it could be generally applied to predict any other global quantity of the halo, such as its concentration, spin, characteristic age, etc. Depending on the specific global quantity considered, different galaxy or subhalo features may be required in order to maximize accuracy, which in principle could differ from those employed here. Moreover, one could apply GNN architectures to other problems, such as edge prediction. For instance, given a set of galaxies, a GNN could be trained to identify those which conform separate halos, as a ML alternative to friend-of-friends algorithms. Furthermore, note that our GNNs are trained to marginalize over baryonic feedback and cosmological parameters, but they cannot identify the specific parameters underlying the simulation of a given halo. However, a GNN can be explicitly trained to infer the values of the cosmological parameters from a simulation employing its galactic population, a task which has been tackled with CAMELS catalogues in \cite{2022arXiv220413713V}.

It would be desirable to synthetize the predictive power of the GNNs into an analytical formula, via symbolic regression, as has been done already in other cosmological contexts \citep{2019arXiv190905862C, Cranmer:2020wew, Shao:2021qoa, Wadekar_2020}. The analytical formulae derived in this way would depend upon the messages exchanged between the nodes. Nonetheless, obtaining a model susceptible to be replicated by a concise formula presents further difficulties. Performing symbolic regression requires interpretable models. One way to achieve this is to enforce sparse weights, obtained, e.g., by applying L1 regularization, which can affect the accuracy of the network. Sometimes, increasing the accuracy of a network reduces its extrapolation properties, and the other way around \citep{Shao:2021qoa}. Hence, deriving simple analytical formulae via symbolic regression from these GNNs remains as a desirable but challenging step for future work.

Our mass inference method based on GNNs is developed for galactic systems composed by central and satellite galaxies, but it could also be applied to galaxy clusters, given that the data description as graphs would be equivalent. While we believe our method will be perfect to perform mass estimates of galaxy clusters, the CAMELS simulations may not be the best training dataset given the small volume they cover, which translates into a lack of galaxy clusters, as already pointed out by previous works \cite{2022arXiv220101305W}.

Besides galaxy clusters, the description as graph-structured data makes this kind of network suitable for other types of astrophysical systems, which are characterized by point clouds. Examples of these may include globular clusters, stellar populations within a galaxy or even  planetary systems. Any point distribution could take advantage of the graph structure presented here for halos, in order to derive global quantities of such systems. An unexplored window remains open to apply all the power of GNNs to astrophysics. 

Given that halo masses can be accurately predicted from numerical simulations by using the method shown here, the natural step forward would be applying this kind of model to real data. Once a GNN model is trained in simulations, one can predict the mass of a real halo using the observed kinematic and internal features of some real galactic systems as input data. Note that this procedure presents further challenges. Spectroscopic measurements would be required to infer the velocity, redshift information and 3D positions of every satellite and central galaxy. The satellite population of distant halos could not be completely observable given their faintness, which could induce some biases in the results. Moreover, additional problems may arise when considering several redshifts at once, such as dealing with redshift space distortions. However, our local galactic neighborhood provides us with a simple scenario to apply our method, where the previous difficulties can be handled. This task is carried out in a companion paper \citep{GNN_MW_M31}, where GNNs trained with CAMELS simulations are employed to predict the masses of the halos containing the Milky Way and Andromeda. In that article, independent predictions for both halos are presented, which are consistent with other standard methods for estimating the dynamical masses of our galaxy and its companion. This represents a success of applying ML models trained with numerical simulations on real data, illustrating the power of artificial intelligence to enhance our knowledge of the Universe.

\section*{Data availability}

The models and implementation of GNNs in PyTorch Geometric, {\tt HaloGraphNet}, are available on  \href{https://github.com/PabloVD/HaloGraphNet}{GitHub \faicon{github}}\footnote{\url{https://github.com/PabloVD/HaloGraphNet}} \citep{HaloGraphNet}. Details on the CAMELS simulations can be found in \url{https://www.camel-simulations.org}.

\section*{Acknowledgements}  

We thank Miles Cranmer for enlightening discussions at early stages of this work. The work of PVD is supported by CIDEGENT/2018/019, CPI-21-108. DAA was supported in part by NSF grants AST-2009687 and AST-2108944, and by the Flatiron Institute, which is supported by the Simons Foundation. The training of the GNNs has been carried out using GPUs from the Tiger cluster at the Princeton University.

\bibliography{bibliography}

\end{document}